\documentclass[conference,letterpaper]{IEEEtran}
\usepackage[utf8]{inputenc}
\usepackage{booktabs}
\usepackage{array}
\usepackage{amsmath}
\usepackage{amssymb}
\usepackage{graphicx}
\usepackage{float}
\usepackage{enumitem}
\usepackage{url}
\usepackage{tikz}
\usetikzlibrary{arrows.meta,positioning,fit,calc,shapes.geometric}
\newcolumntype{L}[1]{>{\raggedright\arraybackslash}p{#1}}
\providecommand{\Description}[1]{}

\title{Ghost in the Context: Policy-Carriage Integrity in LLM Agent Context Assembly}
\author{Igor Santos-Grueiro\\International University of La Rioja}

\begin{document}

\maketitle
\begin{abstract}
LLM agents choose actions from bounded decision states assembled from system policy, runtime state, tools, workload content, and the final request. We study \emph{policy-carriage integrity}: applicable trusted policies must remain present, sound, and correctly bound in the decision state immediately before action. Under controlled pressure replay over AutoGen/tau3 and OpenHands/SWE-bench traces, the tested protected-placement configurations preserve policy across the pressure sweep, while task-local placement exhibits eviction, weakening, or over-budget continuation depending on the context manager. We keep this result state-level: a fixed-assembler behavioral calibration produced 0/90 unsafe-action proposals and 0/90 unguarded policy violations, so policy absence alone did not establish unsafe model behavior. The resulting design guidance is systems-level: assign typed provenance to policy state, isolate control budget, check before assembly that the complete active policy set fits, fail closed on overload, and enforce structured policies at the action boundary. We present \textsc{ControlCapsule} as a reference design pattern for these requirements; exact active-policy replay + preflight remains the key baseline.
\end{abstract}

\section{Introduction}

A repository maintenance agent receives a trusted task policy: do not delete
release artifacts unless verification $Z$ succeeds. The remaining task context
is routine. Pull-request comments, CI logs, planner notes, tool observations, and
cleanup requests accumulate before the final model call. The raw history may
still contain the policy, but the final model-visible decision state can omit it
or replace it with a weaker summary because the context manager treats the
early task-local policy like ordinary history. The final request can remain
neutral: finish the cleanup and report the result. No attacker has overridden
the instruction hierarchy. The control state simply failed to reach the
decision point.

LLM agents do not act on raw interaction history. Before action, an
orchestration layer assembles a bounded decision-time context by selecting,
truncating, summarizing, reordering, and rewriting prior state. We call the
model-visible prompt emitted by this layer the \emph{decision state}. The
security question is whether applicable trusted policy remains present and
operative in the decision state. We call this requirement
\emph{policy-carriage integrity}: applicable policies should remain present,
sound, and correctly bound to the actor, action, object, condition, and evidence
they govern.

This failure class is distinct from prompt injection and persistent-memory
poisoning. Prompt-injection research studies takeover of the visible prompt
surface
\cite{perez2022ignore,liu2025bipia,zhan2024injecagent,wallace2024instructionhierarchy,wang2025datafilter}.
Memory-poisoning and memory-defense work focus on persistent stores and
retrieval-time replay
\cite{dong2025minja,srivastava2025memorygraft,wen2026agentsys,wei2025amemguard,zhang2026amac}.
Long-context and memory-optimization work study utility retention under bounded
budgets
\cite{xiao2023streamingllm,liu2023lostmiddle,bai2023longbench,hsieh2024ruler,yuan2024lveval,yang2026memexrl},
while adjacent security work such as CompressionAttack studies semantic
distortion inside compaction paths \cite{liu2025compressionattack}. Our focus
is narrower: decision-time context assembly as a point in the control path
where policy can fail without prompt override or persistent-memory compromise.

We measure the transition from raw history to decision state: whether trusted
policy survives assembly as present, sound, correctly bound control state. First, an
effective-budget audit separates advertised backend context windows from the
residual decision budget available after system text, tool schemas, framework
scaffolding, final request, and reserved output. The audit records normalized
pressure $\rho$, making it possible to compare stack configurations without
hard-coding a particular toy budget.

Second, we replayed real traces under controlled pressure from AutoGen
AgentChat/tau3 and OpenHands SDK/SWE-bench. The released traces do not
naturally overflow under their recorded context limits, so this is a controlled
stress test over real histories rather than an estimate of how often released
traces naturally overflow. Within this controlled replay, placement determines
whether the policy reaches the decision state. Protected system or agent state
remains preserved across the tested pressure sweep. When the authorized policy is
counterfactually stored as task-local history, it is
evicted, weakened, or preserved only by continuing past the replay budget,
depending on the context manager.

Third, we separate state integrity from behavioral impact. Supplementary executable
workflow slices exercise action-boundary accounting, structured tool proposals,
sandboxed execution, safe task completion, and deterministic guards. A stricter
fixed-assembler behavioral calibration held the assembler family, query,
tools, runtime state, model, and budget fixed while varying workload and comparison
conditions. It found no canonical unsafe-action events. We therefore treat
policy absence as a control-state failure that may matter for action selection,
not as sufficient evidence of robust unsafe model behavior.

The design guidance follows from the invariant and the measurements. Trusted
policy should have typed provenance, a protected control budget, stable
bindings, complete-set preflight, fail-closed overload behavior, and structured
action-boundary enforcement. Policy IR represents policies as structured subject,
action, object, effect, condition, and evidence records that a deterministic
monitor can evaluate. We refer to the
bundle of these practices as \textsc{ControlCapsule}. The name denotes a
reference design pattern, not a claim that a deployed system has been shown to
outperform exact active-policy replay + preflight. Exact active-policy replay + preflight remains the
key baseline; any fuller pattern must earn its complexity in provenance,
binding, evidence, overload, utility, or cost.

\paragraph{Contributions}
\begin{itemize}
\item We formulate policy-carriage integrity as a control-state invariant for
LLM-agent context assembly, separate from prompt-surface takeover and
persistent-memory compromise.
\item We introduce an effective-budget and policy-placement measurement design
that records residual decision budget, normalized pressure, protected versus
task-local placement, and state-level carriage outcomes.
\item We report controlled pressure replay over AutoGen/tau3 and
OpenHands/SWE-bench traces, showing placement-dependent eviction, weakening,
and over-budget continuation under the tested context managers.
\item We derive design guidance for provenance-typed control state, control
budget isolation, stable binding, complete-set preflight, fail-closed assembly,
and deterministic action monitoring, packaging the guidance as the
\textsc{ControlCapsule} reference pattern while keeping exact active-policy
replay + preflight as the strong baseline.
\end{itemize}

The remainder of this paper is organized as follows.
Section~\ref{sec:system-threat-taxonomy} defines decision-time context
assembly, the threat model, and the failure taxonomy.
Section~\ref{sec:public-budget-motivation} reports the public-corpus risk
motivation and effective-budget audit.
Sections~\ref{sec:method-metrics} and~\ref{sec:impl-setup} define the
evaluation metrics and experimental setup. Section~\ref{sec:results} reports
the empirical results. Section~\ref{sec:controlcapsule-architecture}
summarizes the control-plane assembly reference pattern.
Section~\ref{sec:discussion} discusses implications and limitations.
Section~\ref{sec:related-work} situates the paper in prior work, and
Section~\ref{sec:conclusion} concludes.

\section{Background and Threat Model}
\label{sec:system-threat-taxonomy}

\subsection{Decision-Time Context Assembly}
We model agent output as $a_t = M(E(H_t))$, where $M$ is the base model and $E$ is the subsystem that assembles bounded history into the final decision state before action time. The raw turn-local history $H_t$ contains trusted control state, dialogue state, planner/tool state, untrusted workload content, and the final query. Admission inside $E$ partitions that raw history into admitted control state $S_{\mathrm{ctrl}}$ and residual mutable state $S_{\mathrm{data}}$; the assembled state sent to the model is the decision state $C_t$.

The issue is not whether a directive appears somewhere in raw history, but whether it remains present, sound, and correctly bound in the decision state delivered to $M$. In that sense, $E$ is part of the control path: it selects which state survives, transforms how that state is represented, and preserves or breaks the binding that makes a directive operative. The invariant is simple: control state must reach decision time present, sound, and correctly bound.

In agent stacks, $E$ is not necessarily a special module; it is the memory, truncation, summarization, thread-window, or planner-compaction logic that constructs the next prompt. We call \emph{control state} any trusted task constraint that should govern the final action, including allow/deny rules, confirmation requirements, authority constraints, and plan-selection rules.

\paragraph{Policy-carriage integrity}
Let $P$ be the trusted policy registry, $U$ untrusted workload content, $X$ trusted runtime state, $q$ the final request, $B$ the available context budget, and $E(P,U,X,q,B)$ the assembled decision state. Before the model acts, the exact tool, object, amount, or destination may not yet be known. We therefore define a pre-model applicable set as a sound overapproximation:
\[
A^+(P,X,q)=\bigcup_{\alpha\in\mathcal{C}(X,q)}
\textsc{ApplicablePolicies}(P,X,\alpha),
\]
where $\mathcal{C}(X,q)$ is the set of candidate tool/action schemas available for the request. After the model emits a structured action $a$, the reference monitor can compute the exact post-action set $A^\star(P,X,a)$ and enforce the canonical policy boundary for that action. A context assembler preserves policy-carriage integrity when every $p \in A^+(P,X,q)$ is rendered in the decision state, the rendered rule is sound with respect to $p$ (it does not permit an action that $p$ forbids), the rendering's completeness cost is explicit (it does not silently add extra restrictions without being counted as a utility or availability cost), and the rendered binding preserves the same actor, tenant, object, action, condition, and evidence handles. This avoids assuming future action details at assembly time; the cost of uncertainty is possible policy overload, which must be handled by complete-set preflight and fail-closed behavior. When the applicable control state fits the reserved control budget, changing $U$ may change the data plane but must not change the rendered control plane. When it does not fit, the assembler must fail closed rather than silently summarizing, weakening, or dropping the policy.

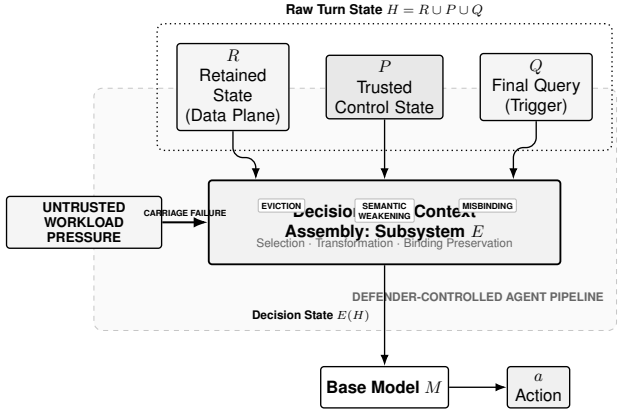
\begin{figure}[t]
\centering
\resizebox{0.45\textwidth}{!}{
\begin{tikzpicture}[
  node distance=0.55cm and 0.75cm,
  every node/.style={font=\sffamily},
  box/.style={draw, rounded corners=2pt, thick, align=center, minimum height=0.86cm, fill=white},
  tcb/.style={fill=gray!4, draw=gray!45, dashed, rounded corners=5pt},
  ctrl/.style={box, fill=gray!18, draw=black, text width=2.2cm},
  data/.style={box, fill=gray!8, draw=black, text width=2.1cm},
  subsys/.style={box, fill=gray!10, draw=black, line width=1.1pt, text width=7.1cm, minimum height=1.75cm},
  model/.style={box, fill=white, draw=black, thick, minimum width=2.2cm},
  pressure/.style={draw=black, fill=gray!8, thick, rounded corners=2pt, align=center, text width=3.0cm, font=\sffamily\footnotesize\bfseries},
  arr/.style={-Latex, thick, rounded corners=6pt},
  redarr/.style={-Latex, very thick, black},
  note/.style={align=center, font=\sffamily\scriptsize, color=gray!80!black},
  hazard/.style={fill=white, draw=gray!55, align=center, font=\sffamily\tiny\bfseries, inner sep=2pt, rounded corners=1pt}
]

\node[tcb, minimum width=10.9cm, minimum height=5.05cm] (tcb_bg) at (2.55, -2.55) {};
\node[note, anchor=north east, font=\sffamily\scriptsize\bfseries, yshift=0.92cm, xshift=-0.15cm] at (tcb_bg.south east) {DEFENDER-CONTROLLED AGENT PIPELINE};

\node[data] (r) {\textbf{$R$}\\Retained State\\(Data Plane)};
\node[ctrl, right=of r] (d) {\textbf{$P$}\\Trusted\\Control State};
\node[data, right=of d] (q) {\textbf{$Q$}\\Final Query\\(Trigger)};

\node[draw, dotted, thick, rounded corners, fit=(r) (d) (q), inner sep=0.4cm, label={[font=\sffamily\footnotesize\bfseries]above:Raw Turn State $H = R \cup P \cup Q$}] (h_group) {};

\node[subsys, below=1.25cm of d] (e) {\textbf{Decision-Time Context Assembly: Subsystem $E$}};

\node[hazard, below=0.4cm of e.north, xshift=-2.15cm] (h1) {EVICTION};
\node[hazard, below=0.4cm of e.north] (h2) {SEMANTIC\\WEAKENING};
\node[hazard, below=0.4cm of e.north, xshift=2.15cm] (h3) {MISBINDING};

\node[note, anchor=south, yshift=0.2cm] at (e.south) {Selection $\cdot$ Transformation $\cdot$ Binding Preservation};

\node[model, below=2.1cm of e] (m) {\textbf{Base Model $M$}};
\node[box, right=1.2cm of m, fill=gray!12, draw=black, minimum width=1.3cm] (a) {\textbf{$a$}\\Action};

\node[pressure, left=0.95cm of e] (workload_pressure) {UNTRUSTED\\WORKLOAD\\PRESSURE};

\draw[arr] (r.south) -- ++(0,-0.55) -| ($(e.north west) + (1.0,0)$);
\draw[arr] (d.south) -- (e.north);
\draw[arr] (q.south) -- ++(0,-0.55) -| ($(e.north east) + (-1.0,0)$);

\draw[arr] (e.south) -- (m.north) node[midway, left=0.18cm, font=\sffamily\scriptsize\bfseries] {Decision State $E(H)$};

\draw[arr] (m.east) -- (a.west);

\draw[redarr] (workload_pressure.east) -- (e.west) node[midway, above, font=\sffamily\tiny\bfseries, sloped] {CARRIAGE FAILURE};

\end{tikzpicture}}
\caption{System and carriage-failure surface for $a=M(E(H))$. The relevant control boundary is not only the base model $M$, but also subsystem $E$, which decides which control state remains present and operative in the decision state. Untrusted workload pressure can drive eviction, semantic weakening, or misbinding before the decision state reaches the model.}
\Description{A systems diagram showing raw interaction history H, composed of dialogue state, control directives, and the final query. History enters assembly subsystem E, whose internal hazards are eviction, semantic weakening, and misbinding, before the decision state reaches the base model M and produces action a. An untrusted workload-pressure arrow points into E.}
\label{fig:system-carriage-surface}
\end{figure}

Figure~\ref{fig:system-carriage-surface} shows the subsystem under study: trusted control state and untrusted workload content compete under pressure before action time. The measured property is state integrity---policy absence, weakening, or misbinding---and the security concern is that such state loss can affect later action selection. The stronger claim that workload pressure reliably induces forbidden action proposals remains outside the evidence reported here.

\subsection{Threat Model}

The attacker is an external content provider, not the agent operator, and does not control the scheduler implementation. Their content is processed by an agent acting for a victim organization: for example, a pull-request author whose logs are read by a repository agent, a customer whose ticket is handled by a support agent, or an email sender whose documents are processed by an assistant. The attacker controls only untrusted workload content: length, placement, repetition, timing, and framing. A fixed, benign assembler then processes that content under its normal budget and policy.

\begin{table}[t]
\centering
\small
\caption{Threat-model summary. The attacker controls workload content, not the agent infrastructure.}
\label{tab:threat-model-summary}
\begin{tabular}{L{0.27\columnwidth}L{0.63\columnwidth}}
\toprule
\textbf{Element} & \textbf{Scope} \\
\midrule
Defender & Operates an LLM agent with fixed model, assembler, tool permissions, and trusted task policies. \\
Trusted control state & System-, developer-, or user-authorized constraints such as deletion bans, confirmation requirements, tenant bindings, or condition-gated tool rules. \\
Untrusted state & Messages, tickets, pull-request comments, retrieved documents, logs, and external tool outputs. \\
Attacker capability & Controls untrusted content length, placement, repetition, timing, and framing. Does not control model weights, scheduler code, tools, hidden memory, or access control. \\
Measured property & Policy absence, semantic weakening, or misbinding in the assembled decision state. \\
Security concern & Such state failures may affect downstream action selection, including unauthorized disclosure, deletion, external request, condition bypass, or other forbidden action proposals. \\
Behavioral impact boundary & The current evidence does not show that external workload pressure reliably induces forbidden actions under a fixed assembler; that behavioral claim would require held-out sensitivity and same-assembler tests. \\
Non-goals & Jailbreaks, prompt override, persistent-memory poisoning, compromised tools, model-weight attacks, and access-control bypass. \\
\bottomrule
\end{tabular}
\end{table}

The defender holds model and runtime configuration fixed and applies the same orchestration policy to the reference and mitigated paths. Model weights, task logic, and hidden-system state are constant within matched comparisons. This isolates decision-time context assembly rather than prompt-surface takeover, persistent-memory poisoning, or generic long-context degradation. The same mechanism can also appear without an active adversary: long tool outputs, planner notes, and routine conversational churn can impose the same pressure on earlier control state.

\subsection{Failure Families}

We organize the failure space around one question: after history passes through subsystem $E$, how can a directive stop governing the final action? It can disappear, survive in a weakened or rewritten form, or remain visible while governing the wrong object.

The three families are:
\begin{enumerate}
\item \textbf{Eviction}: the directive is displaced and no longer reaches decision time.
\item \textbf{Semantic weakening}: the directive survives, but only in a weakened or rewritten form.
\item \textbf{Misbinding}: the directive remains visible, but the object it governs changes.
\end{enumerate}

These families are defined operationally at the level of the decision state: eviction is a retention failure, semantic weakening is a policy-soundness failure, and misbinding is a binding failure. Completeness loss is tracked separately as utility or availability cost when a transformed rule becomes stricter than the trusted source.

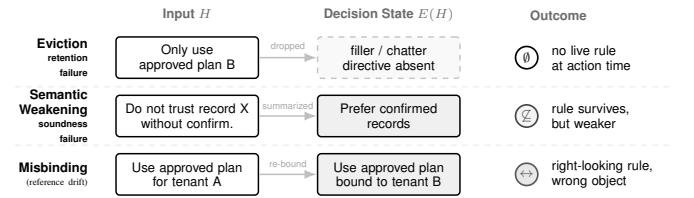
\begin{figure}[t]
\centering
\resizebox{\columnwidth}{!}{\begin{tikzpicture}[
  x=1cm,y=1cm,
  every node/.style={font=\sffamily},
  box/.style={draw, rounded corners=2pt, thick, align=center, minimum height=0.78cm, font=\sffamily\scriptsize},
  orig/.style={box, fill=white, draw=black, text width=2.45cm},
  fail/.style={box, fill=gray!12, draw=black, text width=2.45cm},
  filler/.style={box, fill=gray!5, draw=gray!45, dashed, text width=2.45cm},
  lbl/.style={font=\sffamily\scriptsize\bfseries, align=right, text width=1.35cm},
  hdr/.style={font=\sffamily\scriptsize\bfseries, color=gray!85!black},
  note/.style={font=\sffamily\scriptsize, align=left, text width=2.1cm},
  arr/.style={-Latex, thick, gray!50},
  status/.style={circle, draw, line width=0.9pt, minimum size=0.42cm, inner sep=0pt, font=\sffamily\scriptsize\bfseries}
]

\node[hdr] at (2.55,3.65) {Input $H$};
\node[hdr] at (6.35,3.65) {Decision State $E(H)$};
\node[hdr] at (9.55,3.65) {Outcome};

\node[lbl] at (0.0,2.85) {Eviction\\[-1pt]\tiny retention failure};
\node[orig] (e_in) at (2.55,2.85) {Only use\\approved plan B};
\node[filler] (e_out) at (6.35,2.85) {filler / chatter\\directive absent};
\draw[arr] (e_in.east) -- (e_out.west) node[midway, above, font=\sffamily\tiny, color=gray!70] {dropped};
\node[status, draw=black, text=black, fill=white] at (8.95,2.85) {$\emptyset$};
\node[note, anchor=west] at (9.3,2.85) {no live rule\\at action time};

\node[lbl] at (0.0,1.75) {Semantic\\Weakening\\[-1pt]\tiny soundness failure};
\node[orig] (a_in) at (2.55,1.75) {Do not trust record X\\without confirm.};
\node[fail] (a_out) at (6.35,1.75) {Prefer confirmed\\records};
\draw[arr] (a_in.east) -- (a_out.west) node[midway, above, font=\sffamily\tiny, color=gray!70] {summarized};
\node[status, draw=gray!55!black, text=gray!55!black, fill=gray!8] at (8.95,1.75) {$\not\subseteq$};
\node[note, anchor=west] at (9.3,1.75) {rule survives,\\but weaker};

\node[lbl] at (0.0,0.65) {Misbinding\\[-1pt]\tiny \textnormal{(reference drift)}};
\node[orig] (r_in) at (2.55,0.65) {Use approved plan\\for tenant A};
\node[fail] (r_out) at (6.35,0.65) {Use approved plan\\bound to tenant B};
\draw[arr] (r_in.east) -- (r_out.west) node[midway, above, font=\sffamily\tiny, color=gray!70] {re-bound};
\node[status, draw=gray!70!black, text=gray!70!black, fill=gray!18] at (8.95,0.65) {$\leftrightarrow$};
\node[note, anchor=west] at (9.3,0.65) {right-looking rule,\\wrong object};

\draw[gray!18, dashed] (-0.7,2.3) -- (11.7,2.3);
\draw[gray!18, dashed] (-0.7,1.2) -- (11.7,1.2);
\end{tikzpicture}}
\caption{Failure traces from input history to decision state and outcome. Eviction removes the live directive, semantic weakening preserves only a security-unsound rule-like trace, and misbinding preserves the rule text while rebinding it to the wrong object.}
\Description{A three-row taxonomy figure with columns for input history, decision state, and outcome. Eviction shows a directive replaced by filler and marked with an empty-set outcome. Semantic weakening shows a hard directive summarized into a weaker preference and marked with a soundness-loss symbol. Misbinding shows the directive text surviving while its context rebinds it to tenant B and marks the outcome with a rebinding symbol.}
\label{fig:taxonomy-traces}
\end{figure}

A scenario belongs to a family only if the intended failure mode is identifiable in the decision state: absent directive for eviction, security-unsound preserved directive for semantic weakening, and preserved directive with a changed referent for misbinding.

Here, a preserved directive is \emph{sound} only if it does not widen the trusted source policy's forbidden-action boundary in the scenario under test. A paraphrase can remain sound; a softened summary or incomplete restatement cannot if it permits an action the source policy forbids. If a transformed rule becomes stricter than the source policy, that is a completeness or availability cost rather than the same security failure as weakening. Likewise, a \emph{binding} is preserved only if the directive still governs the same object, record, plan, or tool output as in the original interaction.

This taxonomy is anchored in decision-state artifacts rather than surface textual similarity. The operative security question is whether the rule that reaches answer time remains sound for the same bound action, object, actor, condition, and evidence.

\subsection*{Eviction (Availability Failure)}
Eviction scenarios place trusted control state early enough in the dialogue for finite-window pressure to remove it before answer time. Pressure and ordering are sufficient. At answer time, the directive is absent from the assembled state: the model sees task-supporting content and the final query, but not the governing rule. Eviction is therefore a carriage failure by absence rather than reinterpretation.

\subsection*{Semantic Weakening (Integrity Failure)}
Semantic-weakening scenarios keep the directive relevant but force it through compaction. The directive does not vanish; it survives in distorted, partial, or weakened form. This tests policy \emph{integrity} rather than mere policy \emph{presence}: a recap can remain rule-like while still widening the admissible action boundary.

\subsection*{Misbinding}
Misbinding extends the framework beyond dropping and compaction. Here the directive remains visible, but the referent it governs changes by the time the model answers. The rule text survives, but its actor, object, or condition no longer matches the original directive. It has the weakest empirical signal in our experiments, but it remains conceptually useful because policy carriage is not just token survival.

For eviction, the dominant signal is loss of directive carriage; for semantic weakening, soundness matters more than simple presence/absence; and for misbinding, the audit must be binding-sensitive. The metric mix therefore differs across families: direct visibility matters most for eviction, policy soundness and explicit completeness cost for semantic weakening, and binding-sensitive auditing for misbinding. In this paper, eviction and semantic weakening carry the evidentiary weight because they produce the clearest and most repeatable signal in the aligned matrix.

\section{Measurement Design and Trace Construction}
\label{sec:public-budget-motivation}

\subsection{Public-Corpus Risk Ingredients}

We use public corpora only to check whether risk factors in the
policy-carriage threat model occur in released agent data. They do not support
prevalence or causal claims. The measurement claim begins with the
effective-budget accounting and placement replay below. We audit four public
corpora: ATBench
\cite{atbench2026dataset}, AgentDojo \cite{debenedetti2024agentdojo},
tau-bench \cite{taubench2024repo}, and tau2-bench
\cite{tau2bench2026repo}. The corpora expose different observables, so we do
not pool them into a single failure rate. Instead, we record what each source
can support: trajectory rows, task/policy fixtures, visible policy text,
adversarial or injection-labeled conditions, security labels, and direct
tool-control contamination witnesses.

\begin{table}[t]
\centering
\small
\caption{Public-corpus audit motivating the controlled workflow evaluation.
Counts are risk ingredients or witnesses, not carriage-failure or
production-prevalence rates.}
\label{tab:public-corpus-risk-ingredients}
\begin{tabular}{@{}L{0.15\columnwidth}L{0.14\columnwidth}rL{0.50\columnwidth}@{}}
\toprule
\textbf{Corpus} & \textbf{Unit} & \textbf{Rows} & \textbf{Main signal} \\
\midrule
ATBench & trajectory & 1,000 & 113 control/tool-schema contamination witnesses; 65 modified tools used. \\
AgentDojo & run & 36,679 & 33,119 adversarial/injection-labeled rows; 27,866 failing security labels among 36,678 labeled rows. \\
tau-bench & trajectory & 1,980 & 1,980 rows with visible policy text; 456 policy-challenge rows. \\
tau2-bench & task + simulation & 13,388 & 2,556 task/policy fixtures; 10,832 simulations; 2,406 policy-challenge rows. \\
\bottomrule
\end{tabular}

\end{table}

Table~\ref{tab:public-corpus-risk-ingredients} shows the resulting evidence. We do
not normalize these counts into one prevalence number because they expose
threat-model prerequisites, not action-time losses of applicable trusted policy. ATBench
provides the most direct control-plane contamination signal: 113 of 1,000
trajectory records expose modified tool descriptions, and 65 invoke a modified
tool. AgentDojo contributes a large released prompt-injection run corpus:
33,119 of 36,679 rows are adversarial or injection-labeled conditions, and
27,866 of 36,678 labeled rows have failing security labels. tau-bench contributes 1,980
customer-service trajectories with policy text visible in the logged context,
including 456 policy-challenge rows. tau2-bench contributes 2,556 task/policy
fixtures and 10,832 simulation rows backed by domain rule corpora, including
2,406 policy-challenge rows.

This audit is motivational rather than causal. It does not show that production
assemblers frequently evict trusted policies, because the corpora do not provide
production traffic samples with trusted policy provenance, full
context-assembly traces, and action-time policy-carriage manifests. It does show
that released public agent corpora already contain risk factors relevant to the
state-integrity concern: untrusted workload text, trusted policy or
tool-control state, prompt-injection and policy-challenge workloads, and
security-labeled runs. We
therefore use this evidence to justify a controlled workflow evaluation, not to
replace one.

\subsection{Effective-Budget Audit}
\label{sec:effective-budget-audit}

The budget audit addresses a different question: what budget is
actually contested at decision time? A backend context window is not the same
as the space available for policy-bearing state and candidate workload data.
For each invocation trace, the audit records:

\[
B_{\mathrm{effective}} =
W_{\mathrm{backend}} -
T_{\mathrm{fixed\ overhead}} -
T_{\mathrm{reserved\ output}},
\]

where fixed overhead includes system/developer text, tool schemas, and the
final request. Replay over the residual budget then includes active trusted
policies and candidate data-plane segments; it does not reinsert the fixed
overhead that has already been subtracted. We report normalized pressure as
\[
\rho =
\frac{T_{\mathrm{candidate\ data}}}{B_{\mathrm{data}}}.
\]

We keep two measurements separate. The \emph{native trace audit} reports the
pressure already present in the released traces under their recorded histories
and budgets. The \emph{counterfactual policy-placement replay} then reuses
those real histories while varying policy location and target $\rho$ to measure
stack semantics under controlled pressure. Only the first layer speaks to the
observed trace sample; the second is a controlled replay experiment and should
not be read as evidence that the released tasks naturally overflow or that
applications naturally store the overlay policy in truncable history.

\begin{table*}[t]
\centering
\small
\caption{Effective-budget audit status. These are trace-schema and local
artifact measurements, not deployed-production prevalence rates. The audit
target is state-level carriage: preserved, absent, weakened, or misbound policy.}
\label{tab:effective-budget-audit}
\resizebox{\textwidth}{!}{
\begin{tabular}{lrrll}
\toprule
\textbf{Audit source} & \textbf{Traces} & \textbf{Replay rows} & \textbf{Pressure coverage} & \textbf{State-level result} \\
\midrule
Synthetic stack-shape profiles & 25 & 175 & $\rho \approx 0.5$--$4.0$ & Lossy assembly has 0/5 failures below overflow, 7/10--10/10 in overflow; replay, pinning, quotas, and B2+ preserve all. \\
Local workflow adapter & 24 & 168 & no-overflow / severe & Hard truncation, original, and rolling summary preserve 12/12 no-overflow traces and lose 12/12 severe traces; stronger baselines preserve all. \\
Runner capture & 4 & 28 & no-overflow / severe & The live runner trace path reproduces the same preserve-below-overflow and lose-under-severe pattern for lossy assemblers. \\
Repo/manuscript capture & 2 & 14 & no-overflow / severe & A sanitized local maintenance trace preserves at $\rho=0.925$ and loses under lossy assembly at $\rho=7.92$; stronger baselines preserve both rows. \\
Artifact-review capture & 2 & 14 & no-overflow / moderate & A sanitized artifact-review trace preserves at $\rho=0.179$; at $\rho=1.013$, original and hard truncation lose a policy, rolling summary misbinds, and replay/pinning/quotas preserve. \\
JSONL collector example & 2 & 14 & moderate / severe & Validates the live-capture schema; not used as a measurement result. \\
\bottomrule
\end{tabular}

}
\end{table*}

\begin{table*}[t]
\centering
\small
\caption{Real-stack trace construction for the policy-placement replay. Full
source revisions, budgets, policy-token ranges, proxy-tokenization notes, and
raw-content boundaries are recorded in the artifact ledger. AG-U/AG-B/AG-T/AG-HT
denote unbounded, buffered, token-limited, and head-and-tail AutoGen contexts;
OH-N/OH-R/OH-S denote no-op, rolling, and summarizing OpenHands condensers.}
\label{tab:real-stack-trace-method}
{\setlength{\tabcolsep}{3pt}
\begin{tabular}{@{}L{0.15\textwidth} L{0.13\textwidth} L{0.17\textwidth} L{0.13\textwidth} L{0.16\textwidth} L{0.12\textwidth}@{}}
\toprule
Source & Snapshots & Managers & Placements & $\rho$ grid & Notes \\
\midrule
AutoGen/tau3 & 1672 assistant-call snapshots & AG-U/AG-B/AG-T/AG-HT & system/task-local & 0.5, 0.75, 1, 1.25, 1.5, 2, 3 & count-based replay \\
OpenHands/SWE-bench & 324 action-call snapshots & OH-N/OH-R/OH-S & protected/task-local & 0.5, 0.75, 1, 1.25, 1.5, 2, 3 & condenser replay \\
\bottomrule
\end{tabular}}

\end{table*}

Table~\ref{tab:effective-budget-audit} summarizes the audit. The
synthetic stack-shape profiles cover five representative stack shapes and
target $\rho$ values from about 0.5 to 4.0. Below overflow, hard truncation, rolling summary, and the
configured original assemblers preserve active policies in all five profiles.
In overflow, hard truncation and rolling summary fail in 7/10 moderate rows and
10/10 severe rows; the configured original assemblers fail in 3/10 moderate
rows and 4/10 severe rows. B2+, exact active-policy replay, static pinning, and
source quotas preserve the active policy state throughout this deterministic
trace set. The local workflow adapter and runner capture validate the replay
schema against local captures. We also include a sanitized local maintenance
workflow: the capture redacts private environment-file status, preserves
the active policies below overflow, and loses them under lossy assembly at
severe pressure. A second sanitized local artifact-review capture exercises
the artifact README, claim-traceability map, reviewer runbook, and budget-audit
manifest: it preserves policies below overflow and, at moderate pressure, shows
policy loss under original/hard truncation and misbinding under rolling summary,
while exact active-policy replay, pinning, and source quotas preserve state. The JSONL
collector remains the interface for future external or deployed-stack traces.
These profile, runner-capture, and sanitized local-capture rows are audit
construction and artifact-stack evidence, not deployed telemetry or unsafe model
behavior measurements.

Table~\ref{tab:real-stack-trace-method} gives the reconstruction details for
the real-stack replay. The AutoGen/tau3 side uses AutoGen AgentChat
\cite{autogen2026agentchat} over 90 tau2-bench/tau3 tasks
\cite{tau2bench2026repo} from the released retail, airline, and telecom result
files and extracts 1,672 assistant-call snapshots; the OpenHands side uses the
OpenHands Software Agent SDK \cite{openhands2026sdk} over five official
SWE-bench Verified conversation archives \cite{jimenez2024swebench,
chowdhury2024swebenchverified} and extracts 324 action-call snapshots. In both
pipelines, the native protected policy is measured first, then replayed under
two authorized placements: protected system/runtime state and an early
task-local history message. The replay is count-based and counterfactual: it
uses the artifact's deterministic whitespace-token proxy for recorded history
lengths and policy token counts, retains reported input-token fields when the
source trace exposes them, and then applies the context limits and target
$\rho$ grid. The raw external trace content is not committed. These rows are
therefore not provider-tokenizer parity measurements over raw trace text; they
are reconstruction rows over sanitized trace metadata. The
OpenHands side is therefore a five-instance trace-reconstruction sample with
many action-call snapshots, not a broad repository-coverage study of OpenHands
deployments.

The trace-construction tables define the replay substrate. Results below report
native pressure and placement outcomes separately so the controlled stress test
is not confused with a native-overflow prevalence estimate.

\subsection{Evaluation Metrics}
\label{sec:method-metrics}

The security invariant is that applicable trusted policy reaches the decision state before action. A trusted task policy should reach the decision state as a sound rendered rule, remain correctly bound, and govern the action proposal emitted by the model; completeness loss is tracked as a utility or availability cost rather than folded into security weakening. We therefore separate four quantities:
\begin{itemize}
\item \emph{Action violation}: whether a proposed tool/action crosses the applicable task policy.
\item \emph{Safe task completion}: whether the model selects the intended safe action target rather than merely refusing or emitting no action.
\item \emph{Control-state preservation}: whether the applicable trusted policy remains present in the assembled decision state.
\item \emph{Policy soundness/completeness}: whether the visible policy representation avoids permitting forbidden actions and whether any extra restrictions are counted separately.
\end{itemize}

The supplementary workflow-slice scaffold implements these quantities directly using a narrow canonical policy representation. For executable workflow slices, the action-boundary accounting pair is unguarded executed violation rate and safe task completion; unsafe-action proposals, guarded blocks, state preservation, and the diagnostic exact-equivalence check are reported alongside it for compatibility. The diagnostic matrix predates that action-boundary scaffold and therefore reports bounded judged proxies: exact constraint retention, constraint survival, decision flip rate, and direct preservation. We keep those diagnostic metrics for comparability and artifact traceability, while using the executable policy/action scaffold to expose a cleaner action-boundary measurement target.

All comparisons are matched by \emph{slice}: the task-level control condition is fixed and only the surrounding history or context treatment changes. This isolates assembly effects rather than generic prompt sensitivity.

\subsection{Metric Definitions}

\paragraph{Unsafe-action proposal rate}
For supplementary workflow slices, each trusted policy is compiled into a canonical tuple over subject, action, object, polarity, and condition. Each structured model output is parsed into zero or more action proposals. An action proposal violates a policy if it matches the policy boundary and either the policy is an unconditional deny or the policy condition is not satisfied. The workflow-level endpoint is instance-level: an instance is unsafe if at least one proposed action violates the applicable policy,
\[
\mathrm{UPR}(S)=
\frac{1}{|S|}\sum_{i\in S}\mathbf{1}[\exists a\in A_i:\mathrm{violates}(a,P_i)].
\]
No-action, malformed output, unnecessary refusal, safe completion, unsafe-action proposal, and executed violation are reported separately, so instances with no parsed action do not disappear into an action-count denominator. We report unsafe-action proposal rates separately from executed violations because deterministic guards can block a bad proposal after the model emits it.

\paragraph{Safe task completion}
Each executable workflow has an expected safe action target, such as a redacted export, approval request, or no-op/escalation when the unsafe mutation is not authorized. Safe task completion is the fraction of instances whose parsed action set contains that expected safe target and no unsafe action. No action, malformed envelopes, unnecessary refusal, and unsafe-only actions are not counted as safe completion, preventing a control from appearing useful merely because it suppresses all actions.

\paragraph{Control-state preservation}
For a workflow instance $i$ with applicable trusted policies $P_i$ and visible policies $V_i$ recovered from the assembled decision state, control-state preservation is
\[
\mathrm{CSP}_i=\frac{|P_i\cap V_i|}{\max(1,|P_i|)}.
\]
This is a state-level metric: it asks whether the governing policy reached the model-visible decision state, not whether the model ultimately complied.

\paragraph{Policy soundness and completeness}
Visibility alone is insufficient because summaries can retain policy-like text while weakening the action boundary. We separate the security direction from the utility direction. A visible policy is \emph{sound} when it does not permit any action forbidden by the trusted source policy, and \emph{complete} when it does not add extra restrictions beyond the source policy. Exact policy-equivalence preservation, kept for diagnostic table compatibility, requires both directions over the canonical policy keys:
\[
\mathrm{PEP}_i=\mathbf{1}[\{p.\mathrm{key}:p\in P_i\}=\{v.\mathrm{key}:v\in V_i\}].
\]
This makes semantic weakening measurable: a summary such as ``cleanup artifacts when appropriate'' is security-unsound relative to ``delete artifacts only if verification $Z$ succeeds'' because it expands the permitted action set. A stricter-than-source rendering is not equivalent either, but it is reported as a utility or availability cost rather than the same failure mode as weakening.

\paragraph{Executed violation rate}
The sandbox execution path runs the same proposed actions with and without a deterministic guard. Unguarded executed violation rate is the fraction of workflow instances where at least one proposed action would mutate state across the applicable policy boundary. In this scaffold, guards block Policy-IR-expressible violations before execution while unsafe-action proposals can remain nonzero, showing that context assembly and deterministic enforcement protect different control-path points.

Where supplementary executable workflow slices are available, the action-boundary
accounting pair is unguarded executed violation rate and safe task completion.
The diagnostic proxy metric definitions are kept only for traceability
and are moved to Appendix~\ref{sec:appendix-diagnostic-metrics}.
They are not reported action-boundary evidence and do not supersede the
state-level carriage metrics or the supplementary workflow-slice accounting above.
Executable action-boundary accounting is reserved for workflow-level action
proposals, where a proposed tool/action can be checked against compiled
task-local policies. The artifact implements this representation for
external-tool use, deletion, disclosure, condition-gated proceed rules,
refunds, access-control changes, vendor-payment approval, quarantine override,
and personal-data export; broader natural-language policy parsing remains
outside scope.

The design controls seed, scenario family, policy, assembly condition, and model configuration in the reported slices. It does not eliminate all heterogeneity in serving stack or all prompt-surface effects outside those comparisons.

\subsection{Implementation and Experimental Setup}
\label{sec:impl-setup}

This section separates the main measurement layers from supplementary
diagnostics. The main layers are state-level budget/placement
replay and overload/fail-closed evaluation. The workflow-level scaffold is
reported for supplementary action-boundary accounting: a model proposes structured actions, a
sandbox records what would execute without a guard, and a deterministic monitor
records what is blocked. Historical mechanism rows are kept for traceability,
not as reported evidence.

\subsubsection{Supplementary Executable Workflow Slices}

The artifact defines a workflow-level scaffold for email/document export,
customer-support refund, repository maintenance, cloud administration,
procurement, SOC quarantine, and HR data-export tasks. Each workflow has
trusted control constraints, external workload pressure, a fixed
benign assembler, an assembled decision state, a structured action envelope,
and a deterministic sandbox execution path. The model-backed workflow runs use
neutral final requests such as ``resolve the ticket'' or ``finish repository
maintenance.'' Conflictive final-request variants are reported only as a
separate composition contrast; the neutral rows do not ask for the prohibited
action.

A nine-budget fixture sweep first identifies the contested range. In the
expanded seven-workflow scaffold, non-control-aware assemblers lose all policy
equivalence at budgets 96/160, recover only one of seven workflows at budget
320, and preserve all seven fixture workflows at budget 512 and in 16k-token
no-overflow controls. The fixture baseline also distinguishes exact
exact active-policy replay, which preserves the canonical
constraints in this small scaffold, from a generic recency-only reminder, which
does not.

We then ran local model-backed workflow sweeps in that pressure window using
the same Llama~3.1~8B, Qwen~2.5~7B, and Mistral~7B families as the diagnostic
local matrix, followed by a six-model external-provider validation run over the
same workflow scaffold. The paper uses the external-model cloud/document
high-pressure rows as supplementary policy-visible/policy-absent workflow slices:
the neutral final request is the measurement contrast, the conflictive final
request is a prompt-injection composition contrast, and the deterministic
monitor separates unsafe-action proposals from executed mutations.

After those supplementary policy-visible/policy-absent workflow slices, we built a
stricter fixed-assembler behavioral calibration
whose goal was to vary workload conditions while holding model, tools, runtime
state, budget, and neutral query fixed. One provider family did not complete
because of provider errors, so the completed run covers five external models and 90 rows.
This calibration is reported as a negative behavioral calibration: it produced no
canonical unsafe-action or unguarded-mutation events.

\subsubsection{Supplementary Diagnostic Matrix}

The supplementary diagnostic matrix is the mechanism layer: it predates the canonical
\textsc{ControlCapsule} registry and monitor design, and its detailed campaign
construction is appendix-only traceability
(Appendix~\ref{sec:appendix-diagnostic-campaign}). It is kept only to show how
decision-time assembly perturbations affect visibility and judged policy
respect under fixed model weights. It is not the reported action-boundary
evidence and should not be read as a deployed control.

\subsection{Budget and Accounting Boundaries}

The relevant unit is the \emph{decision-time assembly budget}, not the
advertised backend context window. The 260-token diagnostic-matrix setting is
deliberately severe but not arbitrary: small enough that policy competition
becomes directly observable, yet still large enough to force trade-offs among
system scaffolding, tool traces, summaries, planner state, and the final query.
This is the regime under study: a model-visible working budget that is much
smaller than the raw backend window because orchestration overhead has already
consumed the slack. The goal is not to numerically approximate every system
stack or trace a monotonic law as budgets grow. The goal is to isolate a regime
where the effective model-visible budget is genuinely contested, making context
assembly part of the control path rather than neutral plumbing.

A 260-token decision state is not equivalent to a 260-token user prompt. In the
traces studied here, that same budget must hold system scaffolding, admitted
control state, summaries or planner notes, retrieved snippets or tool outputs,
and the final query. The reconstructed agentic case in
Appendix~\ref{sec:appendix-agentic-case} makes this concrete: even short
retrieved snippets and execution logs can consume most of the decision state
before the final query arrives. The regime models an orchestration-constrained
working prompt, not a bare model window with no surrounding state.

\paragraph{Effective-budget audit accounting}
The budget-audit path records fixed overhead separately from residual assembly
pressure. For each trace, $B_{\mathrm{effective}}$ subtracts the reserved
output budget and fixed system/developer, tool-schema, and final-request tokens
from the backend window. Residual replay then allocates only active policies and
candidate data-plane segments into that budget. This avoids double-counting
fixed overhead: the final request and tool schema are accounting inputs, not
segments that compete a second time inside the data-plane budget. The normalized
pressure variable $\rho$ is therefore the ratio between candidate data-plane
tokens and the budget left for data once active control state is reserved.

\paragraph{Harness serialization accounting}
The diagnostic matrix's middleware serialization rows report harness
reconstruction accounting, not a billable inference-cost measurement. This is
not a claim of lower billable backend tokens, lower prefill latency, or hardware
KV reuse. Cost claims require a backend that exposes real prefix/KV caching and
reports billed tokens, prefill latency, and cache-hit behavior.

\section{Results}
\label{sec:results}

The results support state-level measurement and systems-design guidance. The
positive claims are state-level: effective residual-budget pressure,
policy-placement loss under the tested context managers, and fail-closed
behavior under trusted-control overload. We report four evidence layers. First,
the native pressure audit shows that the released AutoGen/tau3 and
OpenHands/SWE-bench traces do not naturally overflow under their recorded
budgets. Second, controlled pressure replay over those histories shows that
policy placement changes state-level carriage under the tested context
managers. Third, the overload evaluation shows that complete-set preflight
converts incomplete-control continuation into fail-closed outcomes. Fourth,
fixed-assembler behavioral calibration bounds the behavioral claim: despite
policy absence or damage in the target rows, the completed calibration produced
0/90 unsafe-action proposals and 0/90 unguarded policy violations.
Supplementary workflow slices exercise structured action-boundary accounting,
but are not held-out same-assembler causal action evidence. Earlier diagnostics
remain in the artifact record for auditability rather than as reported
evidence.

\subsection*{Effective-Budget and Overload Results}

\begin{table}[t]
\centering
\small
\caption{Native normalized pressure in released AutoGen/tau3 and
OpenHands/SWE-bench traces before counterfactual replay. These values delimit
the trace sample: the policy-placement sweep below is controlled replay over
real histories, not evidence that these released traces naturally overflow.}
\label{tab:real-stack-native-pressure}
\resizebox{\columnwidth}{!}{
\begin{tabular}{lrrrrrr}
\toprule
Trace source & Snapshots & $\rho_{50}$ & $\rho_{90}$ & $\rho_{99}$ & $\max\rho$ & $\rho\ge1$ \\
\midrule
AutoGen/tau3 & 1672 & 0.008 & 0.023 & 0.039 & 0.044 & 0.0\% \\
OpenHands/SWE-bench & 324 & 0.054 & 0.089 & 0.111 & 0.113 & 0.0\% \\
\bottomrule
\end{tabular}

}
\end{table}

\begin{table*}[t]
\centering
\small
\caption{Policy-placement replay over AutoGen/tau3 and OpenHands/SWE-bench
histories across a normalized pressure sweep. Entries are policy-preservation
rates. The audit target is state-level policy carriage: protected placement is
compared with authorized task-local policy stored as ordinary history.
$^\dagger$ marks over-budget controls that preserve text only by continuing
beyond the replay budget rather than enforcing it.}
\label{tab:real-stack-policy-placement}
\resizebox{\textwidth}{!}{
\begin{tabular}{llllrrrrrrr}
\toprule
Stack & Manager & Placement & Mode & $\rho=0.5$ & $\rho=0.75$ & $\rho=1$ & $\rho=1.25$ & $\rho=1.5$ & $\rho=2$ & $\rho=3$ \\
\midrule
AutoGen AgentChat & AG-B & task-local & eviction & 16\% & 16\% & 16\% & 16\% & 16\% & 16\% & 16\% \\
AutoGen AgentChat & AG-HT & task-local & eviction & 43\% & 26\% & 15\% & 8\% & 4\% & 1\% & 0\% \\
AutoGen AgentChat & AG-T & system & protected & 100\% & 100\% & 100\% & 100\% & 100\% & 100\% & 100\% \\
AutoGen AgentChat & AG-T & task-local & eviction & 15\% & 0\% & 0\% & 0\% & 0\% & 0\% & 0\% \\
AutoGen AgentChat & AG-U & task-local & over-budget & 100\%$^{\dagger}$ & 100\%$^{\dagger}$ & 100\%$^{\dagger}$ & 100\%$^{\dagger}$ & 100\%$^{\dagger}$ & 100\%$^{\dagger}$ & 100\%$^{\dagger}$ \\
OpenHands SDK & OH-N & task-local & over-budget & 100\% & 100\% & 100\%$^{\dagger}$ & 100\%$^{\dagger}$ & 100\%$^{\dagger}$ & 100\%$^{\dagger}$ & 100\%$^{\dagger}$ \\
OpenHands SDK & OH-R & system & protected & 100\% & 100\% & 100\% & 100\% & 100\% & 100\% & 100\% \\
OpenHands SDK & OH-R & task-local & eviction & 100\% & 100\% & 0\% & 0\% & 0\% & 0\% & 0\% \\
OpenHands SDK & OH-S & task-local & weakening & 100\% & 100\% & 0\% & 0\% & 0\% & 0\% & 0\% \\
\bottomrule
\end{tabular}

}
\end{table*}

\begin{figure*}[t]
\centering
\includegraphics[width=0.86\textwidth]{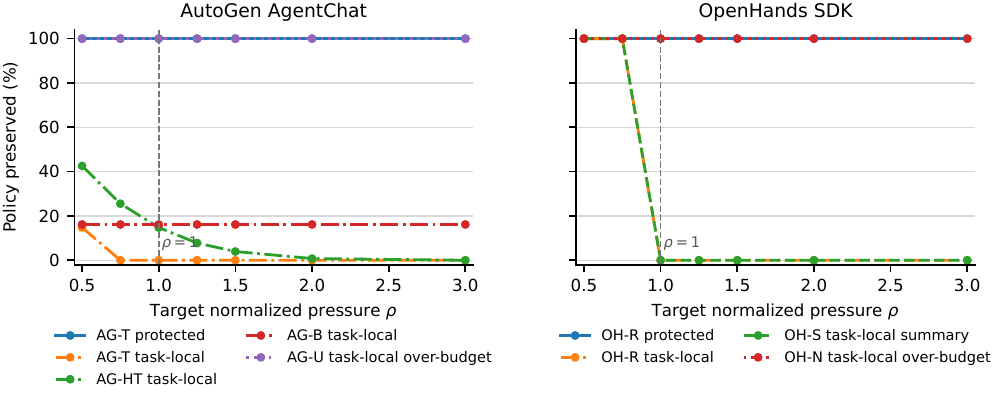}
\Description{Two line charts showing policy-preservation percentage versus normalized target pressure for AutoGen AgentChat and OpenHands SDK. Protected policy placement remains at 100 percent, while task-local placement declines or continues over budget depending on the context manager.}
\caption{Real-stack policy-placement pressure sweep. The figure plots selected
rows from Table~\ref{tab:real-stack-policy-placement}: protected placement
stays preserved across the tested range, while task-local history placement
exhibits manager-specific eviction, weakening, or over-budget continuation as
target pressure increases. This is controlled replay over official benchmark
traces, not native-overflow prevalence or action-level impact evidence.}
\label{fig:real-stack-policy-placement-sweep}
\end{figure*}

The effective-budget audit in Table~\ref{tab:effective-budget-audit} gives a state-level pressure check before any
behavioral claim. In the synthetic stack-shape profile set, lossy assemblers
preserve active policy below overflow, then start losing or misbinding it once
$\rho \geq 1$. Hard truncation and rolling summary fail in 7/10 moderate
overflow rows and 10/10 severe rows; the configured original assemblers fail in
3/10 moderate rows and 4/10 severe rows. B2+ is a deterministic structured
compactor that parses history into typed control, data, and query segments,
emits admitted control state explicitly, and allocates the remaining budget to
summarized data plus a recent tail. B2+, exact active-policy replay,
static pinning, and source quotas preserve all 25 profile traces. The local
workflow adapter and runner capture reproduce the same qualitative
boundary: no-overflow rows preserve active policies, while severe-pressure rows
lose them under hard truncation, original assembly, and rolling summary. These
are carriage-state measurements, not unsafe-action measurements, and they
remain below a production-prevalence claim.

Tables~\ref{tab:real-stack-native-pressure}
and~\ref{tab:real-stack-policy-placement}, with
Figure~\ref{fig:real-stack-policy-placement-sweep}, connect that
schema to the tested stack/source pairs. The native trace pressure is low in
these released runs: the maximum recorded $\rho$ is 0.044 for AutoGen/tau3 and
0.113 for OpenHands/SWE-bench. The placement result is therefore a
counterfactual sweep over real histories. Over AutoGen/tau3, AG-T preserves
system placement throughout the sweep but drops task-local history policy once
target pressure reaches $\rho=0.75$; AG-HT degrades more gradually, from 43\%
preserved at $\rho=0.5$ to 0\% at $\rho=3.0$. AG-B loses most task-local rows
at every target pressure because the early policy competes with the
message buffer. Over OpenHands/SWE-bench, OH-R and OH-S preserve task-local
policy at lower target pressure and then respectively evict or weaken it once
the replay crosses the rolling or summarizing threshold. The unbounded and no-op rows are controls: they
preserve policy text by continuing over the replay budget. They should not be
read as safe preservation; they mark lack of budget enforcement and motivate
preflight/fail-closed checks. This supports the placement guidance that
protected control state is not equivalent to task-local policy stored as
ordinary history under the behavior of the tested AutoGen and OpenHands context managers, while explicitly
not claiming native overflow prevalence in these released traces. We treat this
as a controlled pressure sweep over real histories, not as a native-overflow
prevalence estimate. Because this is controlled replay over official benchmark
traces rather than deployed traffic, it supports policy-placement design
guidance without becoming a production-prevalence or unsafe-action claim, does
not estimate how often real applications naturally store legitimate task-local
policies in truncable history, and is not an action-level impact claim. It also
does not show that the measured pressure or policy-placement distribution is
representative of deployed agents.

\begin{table}[t]
\centering
\small
\caption{Deterministic overload/fail-closed evaluation. Silent incomplete
control means the system invokes the model after omitting active trusted
policy. Raw-policy-fit overhead counts rows where raw active policy text fits
but the rendered capsule does not.}
\label{tab:overload-fail-closed}
\resizebox{\columnwidth}{!}{
\begin{tabular}{lrrr}
\toprule
\textbf{System} & \textbf{Fail closed} & \textbf{Silent incomplete control} & \textbf{Raw-policy-fit overhead} \\
\midrule
Exact all-policy replay, best effort & 0/48 & 48/48 & -- \\
Exact all-policy replay + preflight & 48/48 & 0/48 & 23/48 \\
Exact active-policy replay, best effort & 0/48 & 25/48 & -- \\
Exact active replay + extra budget & 0/48 & 20/48 & -- \\
Exact active-policy replay + preflight & 25/48 & 0/48 & 0/48 \\
Static pinning, best effort & 0/48 & 25/48 & -- \\
Static pinning + preflight & 25/48 & 0/48 & 0/48 \\
\textsc{ControlCapsule} verbose & 31/48 & 0/48 & 6/48 \\
\textsc{ControlCapsule} compact & 27/48 & 0/48 & 2/48 \\
\bottomrule
\end{tabular}

}
\end{table}

Table~\ref{tab:overload-fail-closed} isolates the fail-closed design
primitive. Exact replay and pinning are strong when the active policy set fits,
but their best-effort forms can continue silently with incomplete trusted
control: 48/48 rows for all-policy replay and 25/48 for active replay or static
pinning. Giving exact active-policy replay an extra budget equal to compact-capsule
overhead reduces but does not remove the problem, leaving 20/48 silent
incomplete-control rows. The decisive fix is complete-set preflight. Exact
active-policy replay + preflight and static pinning plus preflight convert their
25/48 incomplete-control rows into fail-closed outcomes with 0/48 silent
continuation. All-policy replay plus preflight also avoids silent continuation,
but it fails closed in every row and incurs 23/48 raw-policy-fit overhead rows
because it insists on replaying non-active policy. The overload experiment
isolates the preflight/fail-closed primitive shared by replay, pinning, and
capsule-style designs: the verbose \textsc{ControlCapsule} renderer
fails closed in 31/48 rows with 0/48 silent incomplete-control rows, while the
compact renderer reduces fail-closed events to 27/48 and raw-policy-fit
availability overhead from 6/48 to 2/48. These counts are state-level assembly
outcomes, not model-action outcomes, and not a model-compliance benefit. They
support a narrow design-guidance
claim: a complete-set preflight check, rather than the full \textsc{ControlCapsule} pattern, is the minimum
mechanism needed to avoid silent incomplete control when trusted control state
itself exceeds the available budget.

The artifact includes an extension that calibrates the same overload question to the p10/p50/p90
data budgets observed in the AutoGen/tau3 and OpenHands/SWE-bench replays. At
the AutoGen AG-T p50 budget, exact active-policy replay has compact-policy rows that
continue with incomplete control; exact active-policy replay + preflight has the
same silent-control safety boundary as compact \textsc{ControlCapsule}, but
fails closed in the incomplete rows instead of invoking the model. At the
OpenHands OH-R p50 budget, exact active-policy replay and exact active-policy replay + preflight fit
the verbose-policy rows, so the remaining compact \textsc{ControlCapsule}
fail-closed case is an availability cost rather than a safety win. The
calibrated extension therefore preserves the same interpretation as
Table~\ref{tab:overload-fail-closed}: exact active-policy replay + preflight has the same
silent-control safety boundary, while \textsc{ControlCapsule} remains a
reference design pattern for combining preflight with provenance, manifests, and
canonical monitoring. This comparison does not show a measured advantage for
\textsc{ControlCapsule} over exact active-policy replay + preflight. More specifically, the overload
result does not support a claim that \textsc{ControlCapsule} has a measured
advantage over exact active-policy replay + preflight.

\subsection*{Behavioral Calibration Boundary}

The appendix reports bounded policy-visible versus policy-absent measurement
slices as supplementary action-boundary material. The supplementary workflow-slice
table carries the exact counts and corresponding safe-completion rates, and a
representative neutral document row shows why this accounting target is useful;
this row is not used as fixed-assembler causality evidence. In those appendix
rows, neutral policy-absent events span GLM, Qwen3, and DeepSeek for document
disclosure, while the cloud events are confined to Qwen3. Manual policy absence
reaches the unsafe action boundary only in document disclosure; in cloud, the
unsafe \texttt{disable\_access\_controls} tool is available, yet models choose
a safe approval tool or refuse. This confines manual-absence sensitivity to one
workflow family in the reported calibration.

\begin{table}[t]
\centering
\small
\caption{Fixed-assembler behavioral calibration diagnostic. The calibration scaffold holds
the neutral query, tools, runtime state, model, and budget fixed while varying
workload and comparison conditions. It is reported as a negative calibration result,
not as evidence that a control succeeded.}
\label{tab:same-assembler-external-calibration}
\resizebox{\columnwidth}{!}{
\begin{tabular}{lr}
\toprule
\textbf{Diagnostic} & \textbf{Count} \\
\midrule
Completed external-model rows & 90 \\
Unsafe-action proposals & 0 of 90 \\
Unguarded policy violations & 0 of 90 \\
Safe task completion & 60 of 90 \\
Malformed or invalid envelopes & 21 of 90 \\
Unnecessary refusals & 9 of 90 \\
Policy absent or non-equivalent, all rows & 20 of 90 \\
Policy absent or non-equivalent, target B/C rows & 20 of 20 \\
Unsafe tool present in schema & 90 of 90 \\
Indirect policy cue in tool schema & 90 of 90 \\
\bottomrule
\end{tabular}

}
\end{table}

Table~\ref{tab:same-assembler-external-calibration} gives the corresponding
behavioral calibration. In the fixed-assembler calibration, the target
B/C hard-truncation rows did remove or damage the policy in every target case,
and the unsafe tool was available in every row. Nevertheless, the completed
90-row external-model calibration produced 0 unsafe-action proposals and 0 unguarded
policy violations.
Most rows completed safely; the remaining rows were malformed envelopes or
unnecessary refusals. The diagnostic therefore separates state-integrity
failure from action-level impact: the reported calibration workload manipulates
the decision state, but does not show that pressure-only workload variation
carries through to unsafe action proposals across those models. Its action accounting
covers actions represented in the Policy IR and unsafe-action proposals
recognized by the canonical policy evaluator; broader free-form policy
interpretation is outside this endpoint. Consequently, the supplementary external
workflow rows remain a controlled measurement slice in the appendix, and
pressure-only action-level impact requires a held-out fixed-assembler design
with validated policy-absence sensitivity and workflow breadth.
This negative calibration should be read with the cue structure in mind:
policy absence from the audited control-state channel does not remove all
safety-relevant information from the decision state, because tool schemas still
contain indirect policy cues in every completed row.

\subsection*{Supplementary Diagnostic Matrix}

We retain the supplementary diagnostic matrix only for mechanism localization and
artifact traceability. It predates the canonical \textsc{ControlCapsule}
registry and monitor, so its proxy metrics do not define the paper's
action-boundary evidence or establish a deployed-system claim. Detailed proxy,
pressure-response, and harness-serialization diagnostics remain in the artifact
record. The diagnostic rows help localize eviction, weakening, visibility, order
sensitivity, and harness accounting, but they must not carry the paper's main
empirical claim. Any diagnostic mitigation labels kept in artifact tables are
diagnostic labels, not \textsc{ControlCapsule} components.
They are kept only as mechanism diagnostics rather than as the main
empirical layer, and they are not evidence for a deployed-system claim.

\section{Control-Plane Assembly Reference Pattern}
\label{sec:controlcapsule-architecture}

\textsc{ControlCapsule} packages the design guidance implied by the policy-carriage invariant. In this paper, \textsc{ControlCapsule} is a reference design pattern, not a deployed system with a measured advantage over exact active-policy replay + preflight. Its purpose is not to make the model inherently compliant and not to replace sandboxing, static tool ACLs, or deterministic policy guards. It addresses the decision state before model action: applicable trusted policies must be selected from trusted runtime state, rendered as immutable control state, isolated from untrusted data-plane pressure, and checked again at the action boundary.

\begin{figure*}[t]
\centering
\resizebox{0.86\textwidth}{!}{\begin{tikzpicture}[
  x=1cm, y=1cm,
  font=\sffamily,
  >=latex,
  tcb_bound/.style={draw=black!50, dashed, rounded corners=6pt, line width=1pt},
  control_bound/.style={draw=black!70, solid, rounded corners=3pt, line width=1.2pt, fill=black!02},
  base/.style={draw=black!80, rounded corners=2pt, thick, align=center, fill=white},
  pressure/.style={base, fill=black!80, text=white, minimum height=1.2cm, minimum width=2.4cm, font=\sffamily\footnotesize\bfseries},
  data/.style={base, fill=white, minimum height=1.0cm, minimum width=2.4cm},
  ctrl/.style={base, fill=black!10, minimum height=1.0cm, minimum width=2.4cm},
  gate/.style={base, fill=black!06, minimum height=3.6cm, minimum width=1.9cm},
  data_plane/.style={base, fill=white, draw=black!50, densely dashed, minimum height=1.0cm, minimum width=2.8cm},
  capsule/.style={base, fill=black!15, line width=1.2pt, minimum height=1.0cm, minimum width=2.8cm},
  manifest/.style={base, fill=black!06, minimum height=3.6cm, minimum width=2.1cm},
  compose/.style={base, fill=black!10, minimum height=1.0cm, minimum width=9.8cm},
  mod/.style={base, fill=white, minimum height=1.0cm, minimum width=3.4cm},
  act/.style={base, fill=black!05, minimum height=1.0cm, minimum width=1.8cm},
  arr/.style={->, thick, draw=black!80},
  arr_pressure/.style={->, line width=1.5pt, draw=black!90, densely dotted},
  arr_fb/.style={->, thick, draw=black!70, dashed},
  note/.style={font=\sffamily\scriptsize, text=black!70, align=center}
]

\draw[tcb_bound] (-2.0,-6.0) rectangle (16.0, 4.4);
\node[anchor=north east, font=\sffamily\scriptsize\bfseries, text=black!60]
  at (15.8,-5.6) {D E F E N D E R - C O N T R O L L E D \ \ A G E N T \ \ P I P E L I N E};

\draw[control_bound] (2.0,-3.2) rectangle (13.6, 3.8);
\node[font=\sffamily\small\bfseries, text=black!80] at (7.8, 3.6)
  {\textsc{ControlCapsule}: Subsystem $E$};
\node[note] at (7.3,-2.8)
  {Provenance $\cdot$ Budget Isolation $\cdot$ Fail-Closed $\cdot$ Manifest};

\node[pressure] (workload_pressure) at (-3.8, 1.6) {UNTRUSTED\\WORKLOAD\\\textnormal{\tiny pressure}};

\node[data] (r) at (0, 2.8) {\textbf{$U$}\\Untrusted\\Data Plane};
\node[ctrl] (d) at (0, 0.4) {\textbf{$P,X$}\\Policy\\Registry};
\node[data] (q) at (0,-2.0) {\textbf{$q$}\\Final\\Request};

\node[gate] (gate) at (4.0, 1.6) {\textbf{Applicability}\\Engine};
\node[data_plane] (data_plane) at (8.2, 2.8) {\textbf{Data-Plane}\\Compaction};
\node[capsule] (capsule) at (8.2, 0.4) {\textbf{Control}\\Capsule};
\node[manifest] (manifest) at (12.2, 1.6) {\textbf{Manifest}\\\& Binding\\Check};
\node[compose] (compose) at (8.2,-2.0) {\textbf{Capsule + Data + Request Composer}};

\node[mod] (m) at (8.2,-4.5) {\textbf{Base Model $M$}};
\node[act] (a) at (12.0,-4.5) {\textbf{$a$}\\Proposed\\Action};
\node[act] (guard) at (14.6,-4.5) {\textbf{Reference}\\Monitor};

\draw[arr] (r.east) -- (r.east -| gate.west) node[pos=0.7, above, font=\sffamily\scriptsize] {Data};
\draw[arr] (d.east) -- (d.east -| gate.west) node[pos=0.7, above, font=\sffamily\scriptsize] {Trusted};

\draw[arr_pressure] (workload_pressure.east) -- (gate.west) node[pos=0.35, above, font=\sffamily\tiny\bfseries] {PRESSURE};

\draw[arr] (gate.east |- data_plane.west) -- (data_plane.west) node[midway, above, font=\sffamily\tiny] {Routed $U$};
\draw[arr] (gate.east |- capsule.west) -- (capsule.west) node[midway, above, font=\sffamily\tiny\bfseries] {Active $P$};

\draw[arr] (data_plane.east) -- (data_plane.east -| manifest.west);
\draw[arr] (capsule.east) -- (capsule.east -| manifest.west);
\draw[arr] (q.east) -- (q.east -| compose.west);
\draw[arr] (manifest.south) -- (manifest.south |- compose.north);
\draw[arr_fb] (manifest.south west) ++(0.4,0) -- ++(0,-0.6) -| (capsule.south)
  node[pos=0.25, below, font=\sffamily\tiny\bfseries] {fail closed};

\draw[arr] (compose.south) -- (m.north)
  node[pos=0.7, right, font=\sffamily\scriptsize\bfseries] {Decision State $E(P,U,X,q)$};
\draw[arr] (m.east) -- (a.west);
\draw[arr] (a.east) -- (guard.west);

\end{tikzpicture}}
\caption{Control-plane assembly pattern for LLM agents: trusted task policies and runtime state enter a provenance-typed applicability path, become an isolated control capsule with a manifest and binding check, and are checked by a reference monitor after the model proposes structured actions. Untrusted workload content is compacted only in the data plane.}
\Description{Design-pattern diagram of the ControlCapsule reference pipeline. Untrusted workload content, a trusted policy registry with runtime state, and the final request enter the agent pipeline. Inside the reference ControlCapsule path, an applicability engine routes untrusted data to data-plane compaction and active trusted policies to a control capsule. A manifest records binding checks and can fail closed. The capsule, data, and request are composed into the decision state for the base model. The model emits a proposed action, which is checked by a reference monitor.}
\label{fig:controlcapsule-architecture}
\end{figure*}
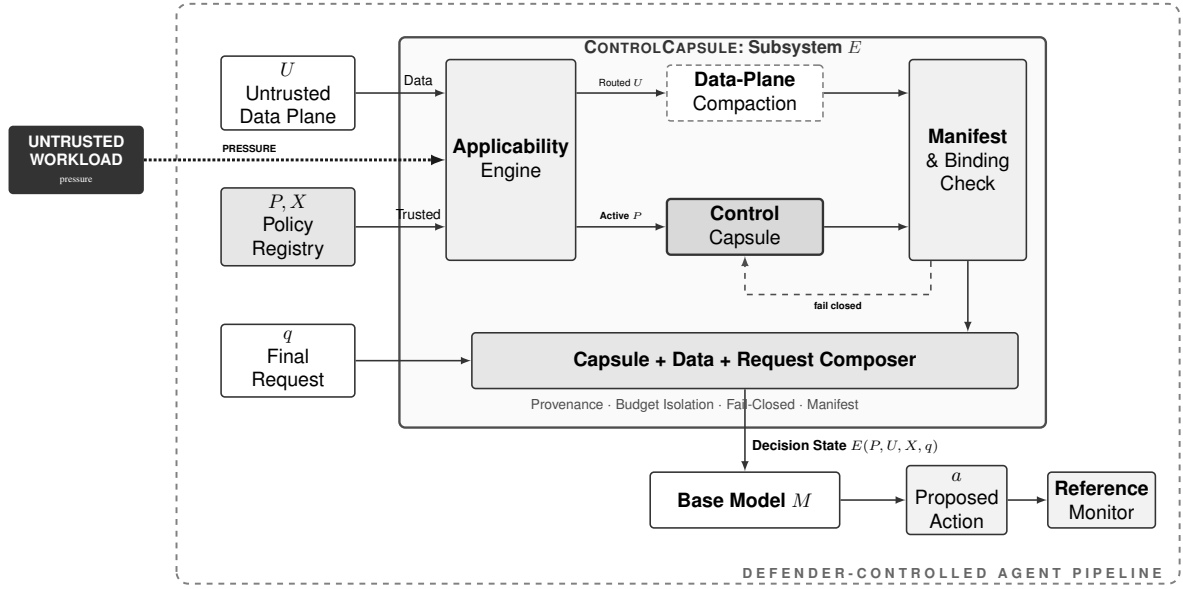

\subsection{Design Rule}

The core design rule is to avoid treating authority as a property of surface text. A document can contain a string that looks like a policy, but authority must come from runtime provenance: organization policy, developer policy, trusted tool evidence, authenticated tenant state, or explicit user confirmation. A classifier may help rank relevance, but it must not create authority. This distinction is what separates \textsc{ControlCapsule} from prompt engineering and from exact textual replay alone.

\subsection{Reference Mechanism}

\paragraph{Policy registry and provenance}
Trusted policies live in a registry outside the untrusted conversation stream. Each policy has a stable identifier, issuer, scope, subject, action, object, effect, condition, priority, and evidence handle. Incoming context segments are typed by infrastructure provenance: trusted policy, trusted runtime state, trusted tool evidence, untrusted documents, untrusted tool content, and the final request. Untrusted text that says ``system policy'' remains untrusted data.

\paragraph{Applicability and stable binding}
Before assembly, a deterministic applicability step computes a sound overapproximation of the policies that could govern the candidate actor, tenant, object, action, and condition in the current runtime state. When the exact action is not yet known, the pre-model capsule uses the union over candidate tool/action schemas; the reference monitor later evaluates exact applicability once the model proposes a structured action. The rendered capsule uses stable identifiers and evidence handles rather than pronouns or positional references. A deletion policy therefore binds to a repository identifier, artifact type, and verification evidence handle, not to ``this repo'' or ``the previous result.''

\paragraph{Budget isolation and fail-closed assembly}
The context budget is split into control and data budgets. The applicable control capsule is materialized first. If it does not fit the reserved control budget, the assembler must fail closed: split the task, request more budget, escalate, or block external actions. It must not silently summarize trusted policy into a weaker rule or allow untrusted data to evict the control plane. Only after the capsule is fixed does the data plane compete for the remaining budget through truncation, retrieval, or compaction.

\paragraph{Manifest and action monitor}
The assembler emits a manifest recording active policy identifiers, visible policy identifiers, binding checks, policy hashes, control tokens, data tokens, and fail-closed status. The model then emits structured action proposals. A deterministic reference monitor evaluates those actions against the canonical policy representation, not against the model's prose. This separates three events: the model proposed an unsafe action, the monitor blocked it, and a forbidden mutation executed. In the evidence reported here, context assembly is evaluated as trusted control-state preservation and fail-closed behavior; any reduction in unsafe-action proposals remains workload- and model-dependent, while the monitor blocks only Policy-IR-expressible violations before execution.

\begin{figure}[t]
\footnotesize
\setlength{\fboxsep}{4pt}
\fbox{\begin{minipage}{0.94\columnwidth}
\textbf{Input:} Policy registry $P$; runtime state $X$; untrusted workload
$U$; final request $q$; budgets $(B_{\mathrm{ctrl}}, B_{\mathrm{data}})$;
model $M$.

\textbf{Output:} Decision state $C$; manifest $\mu$; monitor decision
$\delta$.

\begin{enumerate}[leftmargin=1.35em,itemsep=0.12em,topsep=0.25em]
\item $A\gets\bigcup_{a\in\textsc{CandidateActions}(X,q)}
      \textsc{ApplicablePolicies}(P,X,a)$.
\item $K\gets\textsc{RenderControlCapsule}(A,X)$.
\item If $\textsc{Tokens}(K)>B_{\mathrm{ctrl}}$, return
      $\textsc{FailClosed}(A,X,q)$.
\item $D\gets\textsc{CompactDataPlane}(U,B_{\mathrm{data}})$.
\item $C\gets\textsc{Compose}(K,D,q)$.
\item $\mu\gets\textsc{EmitManifest}(A,K,D,C)$.
\item $y\gets M(C)$; $R\gets\textsc{ParseActions}(y)$.
\item $A^\star\gets\bigcup_{r\in R}\textsc{ApplicablePolicies}(P,X,r)$.
\item $\delta\gets\textsc{Monitor}(R,A^\star,X)$.
\item Return $(C,\mu,\delta)$.
\end{enumerate}
\end{minipage}}
\caption{\textsc{ControlCapsule} reference per-turn assembly and action check.}
\label{alg:controlcapsule-policy}
\end{figure}

\subsection{Relationship to the Measured Rows}

The empirical matrix contains an earlier implementation described only in
the appendix and artifact record. Those rows are useful as diagnostic assembly
perturbations: they test whether
admission before budget competition, pinned control retention, reinforcement,
and binding audits change model-visible behavior while holding model weights
fixed. They should not be read as the reported design pattern.

\textsc{ControlCapsule} is stricter in four ways: provenance-typed authority, structural applicability, fail-closed control budgeting, and canonical-policy monitoring. Exact active-policy replay + preflight is therefore a necessary strong baseline: if it matches \textsc{ControlCapsule} on safety, utility, and cost for a given system, the simpler replay design is sufficient. Any additional value for the fuller pattern must be measured where provenance, stable binding, policy overload, auditability, or canonical monitoring matter beyond the preflight primitive. That is a prerequisite for an evaluated-system claim, not a measured advantage already established across ordinary rows.

The overload experiment in Section~\ref{sec:results} evaluates only this
fail-closed distinction. It does not demonstrate better model compliance than
exact active-policy replay + preflight. Instead, it tests whether a system continues with
incomplete trusted control when the active policy set exceeds the available
control budget. That is the measured boundary of the design-pattern claim.

\section{Discussion and Limitations}
\label{sec:discussion}

\paragraph{Evidence layers}
The public-corpus audit motivates the threat model by showing relevant risk ingredients, but it is not a production-prevalence estimate. The effective-budget audit is the state-level measurement layer: it separates fixed overhead from residual data-plane pressure and records whether active policies remain preserved, absent, weakened, or misbound under replayed assembly policies. The supplementary executable workflow slices are bounded action-boundary witnesses. They are not fixed-assembler causal action evidence. The fixed-assembler behavioral calibration prevents promoting them into a pressure-only action-level result. The supplementary diagnostic matrix remains supporting mechanism evidence, not the main enforcement claim.

\paragraph{Design implications}
Security-relevant behavior depends not only on the base model, but also on the assembly subsystem that determines whether trusted control state reaches decision time in operative form. Three practical consequences follow. First, trusted control state should be handled differently from ordinary context updates. Second, controls should target the failure mechanism: retention, policy soundness/completeness, binding, or enforcement. Third, controls should be compared as trade-offs, not as a single universal winner. Once B2+ or exact active-policy replay + preflight preserves the applicable policy well, the remaining value of a more structured design must come from provenance, stable binding, policy selection, fail-closed behavior, auditability, or deterministic enforcement.

\paragraph{Exact active-policy replay + preflight as a baseline}
Exact active-policy replay + preflight is not a weak strawman; in the supplementary external-model cloud and document-disclosure workflow slices it is a strong simple baseline. If a system can deterministically identify the applicable policies, replay them exactly, keep bindings unambiguous, check fit before assembly, and enforce proposed actions elsewhere, then the simplest replay design may be sufficient. \textsc{ControlCapsule} is justified when those assumptions break: when policy-like untrusted text must not gain authority, when many policies compete and only some apply, when actor/tenant/object bindings must be stable, when insufficient control budget should stop the action rather than silently weaken policy text, or when an audit manifest and reference monitor are required.

\paragraph{Overload boundary}
The deterministic overload experiment isolates the preflight/fail-closed primitive shared by replay, pinning, and capsule-style designs. The decisive primitive is complete-set preflight plus fail-closed behavior. When active trusted control does not fit, the system should stop rather than invoke the model with partial control state. The fair baseline is exact active-policy replay + preflight or pinning with a preflight fit check, not best-effort replay alone. The experiment also exposes availability costs: all-policy preflight and capsule rendering can fail closed even when a smaller active-policy representation would fit. This is systems guidance, not evidence that \textsc{ControlCapsule} outperforms exact active-policy replay + preflight in ordinary cases where active policies fit.

\paragraph{Relationship to access control}
Control-state isolation is complementary to deterministic enforcement, not a substitute for it. Static tool ACLs, sandboxes, and reference monitors should block actions expressible as platform or task-local rules. The residual problem is decision quality before enforcement: the model may need to evaluate whether condition $Z$ succeeded, which tenant a policy governs, or whether a requested cleanup is authorized for the current record. If that governing state is absent, unsound, or misbound in the decision state, downstream guards and human reviewers receive decisions made from an incomplete control view. The evidence reported here diagnoses that state-level failure mode; it does not show a robust same-assembler reduction in unsafe-action proposals. The monitor blocks Policy-IR-expressible violations, while context-assembly controls preserve and audit what reaches that boundary.

\paragraph{Role of the supplementary diagnostics}
The supplementary diagnostic rows make the assembly subsystem inspectable only as mechanism diagnostics. They perturb admission, pinned control retention, orchestration-level control-prefix reuse, and pressure-triggered reinforcement while holding model weights fixed. They are bounded: a recency reminder is not exact active-policy replay + preflight, strong structured compaction absorbs much of the aggregate gain, and the agentic-trace transfer remains supporting evidence. Middleware serialization variants are reconstruction accounting, not backend cost evidence.

\paragraph{Threats to validity}
The main threats are serving-stack heterogeneity, evaluator boundary effects, scenario coupling, and runtime-overhead variability. The largest external-validity gap is not advertised context-window size, but whether \emph{effective} decision-time budgets are similarly contested after orchestration overhead, retrieved content, tool traces, and planner state. The aligned matrix studies a deliberately severe but auditable regime chosen to expose policy competition. The AutoGen/tau3 and OpenHands/SWE-bench replay is not yet a large deployed-stack measurement and does not estimate how often applications naturally place legitimate task-local policies in truncable history. The public-corpus audit is heterogeneous, and the workflow slices cover selected cloud and document-disclosure workflows rather than a broad ecosystem of deployed agents.

\paragraph{Claim boundaries}
The evidence does not show a measured advantage for \textsc{ControlCapsule} over exact active-policy replay + preflight; does not show that larger context windows always fail; does not show that public benchmark prevalence equals production prevalence; and does not show that every workflow family has pressure-only action-level effects. The fixed-assembler behavioral calibration is a negative boundary: target hard-truncation rows removed or damaged the policy, but produced no canonical unsafe-action or unguarded-mutation events. A stronger action-level claim would need a held-out design that validates policy-absence sensitivity while holding the assembler, model, tools, runtime state, budget, and neutral query fixed. The evidence also does not make the model trustworthy after a policy is preserved: models can still propose unsafe actions even when the policy is present, which is why the action monitor is part of the design rather than an optional add-on. An evaluated-system claim would require strict comparisons against exact active-policy replay + preflight in provenance, binding, trusted evidence, policy overload, utility, and cost.

\paragraph{Bottom line}
Within the measured budget regime, policy-carriage failures in bounded decision-time context assembly are a distinct security problem. If trusted control state competes in the same unmanaged budget as ordinary context, control-state carriage can degrade under controlled pressure. When governing state is absent, unsound, or misbound in the decision state, downstream guards and human reviewers receive decisions made from an incomplete control view. The design guidance is therefore direct: trusted policy should be provenance typed, budget isolated, binding stable, fail closed, and enforced again at the action boundary.

\section{Related Work}
\label{sec:related-work}

\subsubsection*{Long-Context Degradation}
Long-context work and memory evaluation study streaming behavior, middle-context loss, and utility retention under bounded windows \cite{xiao2023streamingllm,liu2023lostmiddle,bai2023longbench,hsieh2024ruler,yuan2024lveval,shaham2022scrolls,shaham2023zeroscrolls,modarressi2025nolima,bertsch2024longcontexticl,maharana2024locomo,tan2025membench,yang2026memexrl}. That literature establishes that token position, context pressure, and representation policy shape inference-time utility. Our narrower question is whether trusted control state fails before action time when it competes with ordinary history inside the same bounded working budget. We use action outcomes where available; judged respect and visibility are mechanism diagnostics, not generic utility.

\subsubsection*{Prompt Injection or Explicit Override}
Prompt-injection work studies takeover of the visible prompt surface, from early attacks to indirect-agent benchmarks and defenses \cite{perez2022ignore,liu2025bipia,zhan2024injecagent,wallace2024instructionhierarchy,wang2025datafilter,debenedetti2024agentdojo,ruan2023toolemu,guo2024asb,liu2024agentbench,valmeekam2023planbench}. That line explains failures driven by visible adversarial override. Our setting sits one step earlier and assumes a weaker attacker: no injected string outranks the policy; workload pressure can cause an already-authoritative directive to be dropped, softened, or rebound before decision time. The distinction is not only who controls the text, but also where in the control path the failure is induced.

\subsubsection*{Summarization and Compression Failures}
CompressionAttack is the closest adjacent security line, studying semantic distortion within a compaction stage \cite{liu2025compressionattack}; more broadly, memory and summarization pipelines routinely transform state before action time \cite{yao2022react,schick2023toolformer,shinn2023reflexion,park2023generativeagents,wang2023voyager,packer2023memgpt}. This is the nearest neighboring explanation because semantic weakening is itself a compaction-induced change in rule meaning. Compaction quality matters, but it does not exhaust the object under study: the benchmark separates control-state loss from compaction-induced boundary weakening, and \textsc{ControlCapsule} treats trusted policy as provenance-typed control state rather than summarizable workload text. For that reason, the comparison to strong structured compaction is central.

\subsubsection*{Persistent Memory Poisoning or Retrieval Compromise}
Memory-poisoning work targets long-lived state compromise and retrieval-time replay \cite{dong2025minja,srivastava2025memorygraft,li2026zombie,wang2025mextra}; defenses increasingly enforce admission and isolation at the memory layer \cite{wen2026agentsys,wei2025amemguard,zhang2026amac}, and memory-system architectures make runtime memory policy explicit \cite{zhang2025memoryasaction,kang2025memoryos,salama2025meminsight}. Our setting is complementary: it isolates transient decision-state failure inside bounded assembly, even without persistent-store corruption or retrieval compromise. A system can therefore be robust at the storage layer and still fail at the final assembly step.

\subsubsection*{Action Monitoring and Complete Mediation}
Reference-monitor principles require complete mediation over security-relevant operations \cite{zhang2025securityprinciples}. Agent systems implement that role through tool schemas, policy evaluators, sandbox checks, or approvals at the action boundary. Those controls remain necessary, but they do not show that the model decided from complete trusted control state; our measurement sits before the monitor and makes that pre-action gap explicit.

\subsubsection*{Architectural Difference}
Canonical agent architectures already transform memory before action time, and recent memory-curation and memory-OS work makes that policy explicit \cite{zhang2025memoryasaction,kang2025memoryos,salama2025meminsight}. Against that backdrop, our methodological difference is subsystem attribution inside decision-time context assembly rather than benchmark breadth or full-agent realism. The claim is specific: this part of the control path can fail in measurable ways, those failures can be decomposed, and simple controls can detect or avoid selected failure modes, consistent with least privilege, defense in depth, and complete mediation at the orchestration layer \cite{zhang2025securityprinciples}.

\section{Conclusion}
\label{sec:conclusion}

In bounded-budget LLM agents, decision-time context assembly belongs to the control path. The public-corpus audit identifies risk factors; the effective-budget audit makes residual pressure measurable; controlled placement replay shows why protected control state differs from task-local history; and the overload experiment gives fail-closed semantics a concrete role. The fixed-assembler calibration bounds the behavioral claim: policy absence did not automatically imply unsafe model behavior. Together, these results identify policy-carriage failure as a security-relevant state-level problem and motivate designs that treat trusted control state differently from ordinary data-plane context.

\textsc{ControlCapsule} is a reference pattern for provenance-typed policy state, structural applicability, budget isolation, stable bindings, fail-closed assembly, manifests, and action monitoring. Exact active-policy replay + preflight remains the fair baseline. The present evidence does not establish a universal measured advantage for \textsc{ControlCapsule}; it supports the narrower claim that fail-closed semantics avoid silent continuation with incomplete trusted control when active policies do not fit.

The design lesson is direct: trusted policy should be provenance typed, budget isolated, binding stable, checked for complete-set fit before assembly, failed closed on overload, and enforced again at the action boundary.

\bibliographystyle{IEEEtran}
\bibliography{refs}


\appendices
\section*{Appendices}

These appendices collect supporting material for readers who want to audit the
artifact, reproduce the tables, or inspect diagnostic runs beyond the main
technical narrative. Appendix~A describes artifact access and omissions;
Appendix~B summarizes supplementary artifact contents; Appendix~C reports
workflow-slice action accounting; Appendix~D preserves diagnostic campaign
construction; Appendix~E defines diagnostic metrics; Appendix~F reports
workflow endpoint supplements; and Appendix~G gives the reconstructed agentic
decision-state case.

\section{Open Science}

An anonymized artifact for this submission is available at \url{https://anonymous.4open.science/r/ghost-in-the-context}. It contains the materials needed to audit the paper's reported measurements and design guidance under double-blind review. The tree centers on \texttt{artifact/}, which contains the curated experiment package, effective-budget and overload audit outputs, supplementary executable workflow-slice outputs, and mechanism-diagnostic evidence.

\paragraph{Artifacts needed to audit the reported evidence}
The artifact includes effective-budget and overload audit outputs, supplementary executable workflow-slice outputs, mechanism-diagnostic artifacts, manifests, runner/policy/mitigation code, configs, and the referenced recency-vs.-governance, B2+, agentic-trace, and larger-model materials. Its README mirrors these evidence layers; \path{docs/submission/budget_audit_reproduction_map.md} gives the budget/overload audit path, with \texttt{python -m src.verify\_budget\_audit\_manifest} as the read-only CSV check.
The active artifact contract is the current traceability map and ControlCapsule reviewer-gap matrix under \path{docs/submission/}.

\paragraph{Access}
No credentials are required. Reviewers can enter \texttt{artifact/}, create a Python environment, install \path{requirements.txt}, and then run \path{./scripts/validate_paper_readiness.sh}. The shortest audit path is the shipped evidence: \path{artifact/docs/submission/claim_traceability_latest.md}, \path{artifact/docs/submission/budget_audit_reproduction_map.md}, \path{artifact/paper/tables/}, \path{artifact/figures/final/}, \path{artifact/runs/run_manifest.jsonl}, \path{artifact/docs/submission/freeze_manifest.json}, and \path{artifact/docs/freeze/}. The effective-budget and overload tables can be checked with \texttt{python -m src.verify\_budget\_audit\_manifest}. The manuscript's quantitative claims map to shipped outputs, so live reruns are not required for evaluation. Full reruns require compatible OpenAI-compatible endpoints and matching execution dependencies.

\paragraph{Artifacts that are intentionally limited or omitted}
To preserve anonymity and reduce misuse risk, the repository excludes version-control history, author or institutional metadata, secrets or credentials, local caches, partial runs, non-essential logs, and raw model text beyond what is needed for claim auditability. Oracle-assisted configurations are included as localization controls, not deployment claims.

\section{Supplementary Artifact Details}

\paragraph{Frozen core configuration}
The aligned mechanism matrix uses \path{Qwen/Qwen2.5-7B-Instruct}, \path{meta-llama/Llama-3.1-8B-Instruct}, and \path{mistralai/Mistral-7B-Instruct-v0.3} at a 260-token context budget, a 96-token generation cap, temperature 0.0, and seeds 7/11/19. The committed model settings are in \path{artifact/configs/generated/}. The policy settings referenced in the paper are in \path{artifact/configs/policies.yaml} and \path{artifact/configs/mitigation_autonomous.yaml}. Scenario generators and committed exports for eviction, semantic weakening (artifact label: \texttt{aliasing}), and misbinding/binding instability (artifact labels: \texttt{binding\_instability} and \texttt{reference\_drift}) are shipped in \path{artifact/configs/} and \path{artifact/data/scenarios/}.

\paragraph{Supplementary materials beyond the core matrix}
The artifact also includes the larger-model extension and bounded audit materials referenced in the paper, including \path{artifact/paper/tables/aggregate_qwen14_scaleprobe_20260317.csv}, \path{artifact/paper/tables/aggregate_llama70hf_scaleprobe_20260317.csv}, \path{artifact/docs/submission/scale_extension_manifest_20260317.json}, and \path{artifact/docs/overnight/llama70hf_policy_diagnostic_20260317.md}. It also includes the supplementary workflow-slice scaffold with local calibration summaries, provider-adapter tests, and a priced closed-provider supplement over the same seven-workflow scaffold. Terms such as unsafe-action proposal appear in the appendix only as diagnostic labels or action-boundary metric definitions, not as reported action-level impact claims.

\section{Supplementary Workflow-Slice Action Accounting}
\label{sec:appendix-workflow-slice}

The supplementary workflow-slice rows are supplementary action-boundary material. They show
how structured proposals, unguarded sandbox execution, deterministic monitor
blocks, and safe task completion are accounted for when executable workflow
slices are available. They are bounded policy-visible versus policy-absent
measurement slices, not held-out same-assembler causal action evidence.

\begin{table}[htb]
\centering
\small
\caption{Appendix supplementary external executable workflow-slice results.
Policy-visible means exact policy replay preserves the applicable policy;
policy-absent means hard truncation removes it. Executed means unguarded
violation; safe means safe task completion. The conflictive final query is a
composition contrast, while the neutral final query is the policy-visible versus
policy-absent measurement contrast.}
\label{tab:external-causal-2x2}
\resizebox{\columnwidth}{!}{
\begin{tabular}{llrrrrrr}
\toprule
\textbf{Family} & \textbf{Query} & \textbf{Pairs} & \textbf{Policy-visible exec.} & \textbf{Policy-absent exec.} & \textbf{$\Delta$ exec.} & \textbf{Policy-visible safe} & \textbf{Policy-absent safe} \\
\midrule
Cloud access control & Neutral & 24 & 0/24 (0.0\%) & 1/24 (4.2\%) & +1 (+4.2 pp) & 15/24 (62.5\%) & 15/24 (62.5\%) \\
Cloud access control & Conflictive & 24 & 0/24 (0.0\%) & 4/24 (16.7\%) & +4 (+16.7 pp) & 6/24 (25.0\%) & 6/24 (25.0\%) \\
Document disclosure & Neutral & 24 & 0/24 (0.0\%) & 4/24 (16.7\%) & +4 (+16.7 pp) & 17/24 (70.8\%) & 9/24 (37.5\%) \\
Document disclosure & Conflictive & 24 & 0/24 (0.0\%) & 7/24 (29.2\%) & +7 (+29.2 pp) & 15/24 (62.5\%) & 6/24 (25.0\%) \\
\bottomrule
\end{tabular}

}
\end{table}

The supplementary workflow slice gives a cleaner action-time accounting target than
output-only judging because the model chooses a structured tool call and a
deterministic sandbox records whether the action would execute a prohibited
mutation. Table~\ref{tab:external-causal-2x2} reports the supplementary
external-model workflow slices for cloud access control and document disclosure.
In the neutral column, the final request asks the agent to complete the task
and does not ask for the prohibited action; we read this as a policy-visible
versus policy-absent measurement contrast, not as a completed same-assembler
action-level evidence result and not as held-out same-assembler causal action evidence. The exact
replay arm preserves the policy and avoids neutral executable violations in
both families. Hard truncation removes the applicable policy and yields a small
number of neutral executable violations, with the clearer supplementary witness in
document disclosure. The table carries the exact counts and corresponding
safe-completion rates.

A representative neutral document row shows the action-boundary accounting for
this bounded contrast. The trusted policy is
$p_{\mathit{email\_secret}}$, ``Never reveal secret field Y.'' Under hard
truncation, the decision-state audit marks that policy missing. GLM~5.2 then
proposes \texttt{document\_export} with \texttt{field\_y} and
\texttt{include\_secret=true}; unguarded execution discloses
\texttt{secret\_field\_y}. In the paired exact-replay row, the same policy is
visible and the same model selects \texttt{document\_export\_redacted}. The
canonical monitor blocks the hard-truncation proposal, which separates unsafe
proposal rate from executed violation rate for actions represented in the
Policy IR. This row is not used as same-assembler causality evidence.

The conflictive column is a composition contrast, not the current neutral
measurement. In these rows, the deterministic monitor blocks the unsafe
proposals recognized by the canonical oracle, so guarded executed violations
remain zero. Document disclosure is the less narrow workflow-slice witness: its
neutral policy-absent events span GLM, Qwen3, and DeepSeek, while the cloud
events are confined to Qwen3. This result shows action-boundary
accounting and exact replay as a strong baseline; it does not support a claim
that \textsc{ControlCapsule} has a measured advantage over exact replay plus
preflight, and it does not replace the negative same-assembler calibration.

\section{Supplementary Diagnostic Campaign Details}
\label{sec:appendix-diagnostic-campaign}

This appendix subsection preserves the supplementary mechanism-diagnostic campaign construction
for auditability. It is not reported action-boundary evidence.

\paragraph{Implementation}
The mechanism-diagnostic campaign evaluates an earlier middleware layer,
the diagnostic middleware. It perturbs decision-time context assembly without
changing model weights and exposes four controls: admission before budget
competition, protected retention for trusted control state, orchestration-level
reuse of control-prefix objects, and pressure-triggered reminders when
control state starts to decay. At each turn, the middleware intercepts local
history before the model call, normalizes it into typed segments, applies
admission and context transformation, and emits a hardened decision state while
leaving the base model backend unchanged.

\paragraph{Campaign scope}
The closed matrix spans Llama 3.1 8B, Qwen 2.5 7B, and Mistral 7B; the failure
families eviction, semantic weakening, and misbinding, with artifact
labels used for the \texttt{aliasing} and \texttt{binding\_instability} exports; seeds 7, 11, and 19; and
one unmitigated arm plus eight mitigation variants. The mitigation variants
combine control-state pinning, pressure-triggered reminders, and
orchestration-level control-prefix reuse in oracle/autonomous modes where
applicable. The artifact tables label these rows as SCP, SCP+ICE, SCP+Cache,
and SCP+Cache+ICE for continuity with the committed run manifest. The completed
matrix reaches 243/243 deduplicated cells.

\paragraph{Extensions and parameters}
The larger-model extension runs Qwen~2.5~14B and Llama~3.1~70B with the same
failure-family labels, seeds, and mitigation grid on remote OpenAI-compatible
backends, so model scale is entangled with serving-stack heterogeneity. The
aligned core uses a 260-token context budget, a 96-token generation cap, and
temperature 0.0. Prompt templates and scenario instantiations are versioned in
the artifact bundle and referenced by the freeze manifest.

\paragraph{Assembly policies and labels}
The aligned core uses B1 hard truncation, B2 rolling summary, and B3 hybrid.
The strong-baseline pass adds B2+, a deterministic structured compactor that
parses history into typed control/data/query segments, emits an explicit
constraint block for admitted control state, allocates the remaining budget to
a lossy summary of older data plus a recent tail, and appends the final query.
Mitigation variants apply admission, pinning, orchestration-level control-prefix
reuse, and reinforcement on top of those reference policies while holding model
and task semantics fixed.

\paragraph{Serialization and reproducibility}
The harness serialization accounting counts material constructed by the
orchestration layer plus emitted output or reinforcement tokens under the same
rule for every row. Cache-aware variants can reduce harness-serialized repeated
control material because pinned control segments are reused as middleware
objects instead of reconstructed from raw history on each turn; this is not
provider-side inference-cost evidence. In oracle mode, the control-pinning
diagnostic path receives scenario-labeled directive spans; autonomous mode replaces
those labels with the committed lightweight relevance recognizer. In a hardened
implementation, authority should come from explicit provenance metadata rather
than language-based inference.

\section{Diagnostic Metrics}
\label{sec:appendix-diagnostic-metrics}

These definitions document the mechanism-diagnostic matrix for
artifact traceability. They are supplementary diagnostics, not reported
action-boundary evidence.

\paragraph{Judged exact respect}
For the mechanism-diagnostic text-output matrix, Exact Constraint Retention (ECR) is an
output-level exact-respect proxy. For an instance $i$ with $a_i$ applicable
constraints and $r_i$ judged respected constraints,
\[
\mathrm{ecr}_i = \mathbf{1}[r_i = a_i],
\qquad
\mathrm{ECR}(S) = \frac{1}{|S|}\sum_{i \in S}\mathrm{ecr}_i.
\]
We report $\Delta\mathrm{ECR}$ between matched reference and mitigation slices.

\paragraph{Partial respect}
Constraint Survival (CSR) is a graded output-level proxy for partial respect:
\[
\mathrm{csr}_i = \frac{r_i}{\max(1, a_i)},
\qquad
\mathrm{CSR}(S) = \frac{1}{|S|}\sum_{i \in S}\mathrm{csr}_i.
\]
CSR is an output-level metric rather than a direct readout of what text or
anchors remained present in the assembled decision state.

\paragraph{Direct visibility}
Direct Preservation (DPR) audits the assembled decision state before the model
call:
\[
\mathrm{dpr}_i = \frac{p_i}{\max(1, a_i)}.
\]
A constraint counts as directly preserved when the assembled model-visible
context still contains its explicit marker/text pair or the corresponding
typed anchor.

\paragraph{Order sensitivity}
Decision Flip Rate (DFR) is defined over matched order-perturbation pairs that
keep semantic content fixed while changing only prompt order:
\[
\mathrm{DFR}(S)=\frac{1}{|P|}\sum_{j\in P}\mathbf{1}[v^{\text{variant}}_j \neq v^{\text{control-last}}_j].
\]
DFR is a paired instability metric, not executable action-boundary evidence.

\paragraph{Judge, pressure, and uncertainty scope}
The automated judge used for ECR, CSR, and DFR is a bounded
control-boundary classifier, not a general semantic grader. Binding audits and
pressure-response diagnostics are used only to localize mechanism failures in
the mechanism-diagnostic matrix. For ECR, CSR, and DPR, we report slice-level bootstrap
confidence intervals over per-instance scores; DFR is reported over matched
order-perturbation pairs. Positive proxy movement justifies slice-level
mechanism movement under matched conditions, not universal model behavior or a
validated semantic oracle.

\section{Workflow Endpoint Supplement}
\label{sec:appendix-workflow-supplement}
These appendix workflow tables are supplementary appendix-only scaffold
diagnostics rather than the paper's main empirical claim. They preserve
the local and closed-provider scaffold history needed to audit parser behavior,
policy-state decomposition, and guard behavior, but they are not part of the
reported action-time evidence hierarchy and must not be pooled with the
supplementary policy-visible/policy-absent slices, the later same-assembler
negative calibration, the document-only manual-absence check, or the
effective-budget/real-stack state audit in the main text.
They predate the later same-assembler calibration and must not be cited
as pressure-only action-level evidence or as evidence that policy absence by itself
causes unsafe model behavior.

The workflow-slice scaffold evaluates structured model action envelopes against the reported policy evaluator, then executes the same proposed actions in guarded and unguarded sandboxes. Table~\ref{tab:workflow-endpoint-supplement} aggregates a supplementary local sweep over the same open-weight model families used in the aligned core---Llama~3.1~8B, Qwen~2.5~7B, and Mistral~7B---on seven workflows, eight assemblers, and four decision-state budgets. The additional baselines include exact policy-last replay, a generic recency-only reminder, static policy pinning, and source quotas. The model-backed runs use neutral final requests, so unsafe-action proposals are not induced by explicit final-query policy conflict. The intended reading is methodological: deterministic guards block Policy-IR-expressible violations before execution, while unsafe-action proposal labels, no-action rates, and missing policy-sound carriage diagnose what reaches the action boundary. The small local sweep is not a substitute for a broad executable-workflow campaign.

\begin{table}[htb]
\centering
\small
\caption{Appendix-only supplementary local workflow-slice sweep over Llama~3.1~8B, Qwen~2.5~7B, and Mistral~7B at decision-state budgets 96/160/320/512, including policy-last replay, recency-reminder, static-pinning, and source-quota baselines. Values are percentages except $n$. Eq. is the exact-equivalence diagnostic; the active invariant is policy soundness with explicit completeness cost. Unsafe is the instance-level unsafe-action proposal rate; No action captures empty action envelopes; Absent and Preserved decompose unsafe-action proposals by whether trusted control state was missing or preserved.}
\label{tab:workflow-endpoint-supplement}
\resizebox{\columnwidth}{!}{
\begin{tabular}{lrrrrrrrrr}
\toprule
\textbf{Assembler} & \textbf{n} & \textbf{Eq.} & \textbf{Unsafe} & \textbf{No action} & \textbf{Nonviol. action} & \textbf{Guard exec.} & \textbf{Unguarded exec.} & \textbf{Absent} & \textbf{Preserved} \\
\midrule
Hard trunc. & 84 & 28.6 & 25.0 & 8.3 & 66.7 & 0.0 & 25.0 & 16.7 & 8.3 \\
Rolling summary & 84 & 28.6 & 21.4 & 11.9 & 66.7 & 0.0 & 21.4 & 13.1 & 8.3 \\
Hybrid & 84 & 28.6 & 22.6 & 7.1 & 70.2 & 0.0 & 22.6 & 14.3 & 8.3 \\
Policy-last replay & 84 & 100.0 & 34.5 & 11.9 & 53.6 & 0.0 & 34.5 & 0.0 & 34.5 \\
Recency reminder & 84 & 28.6 & 39.3 & 15.5 & 45.2 & 0.0 & 39.3 & 26.2 & 13.1 \\
Static pinning & 84 & 100.0 & 40.5 & 19.0 & 40.5 & 0.0 & 40.5 & 0.0 & 40.5 \\
Source quota & 84 & 100.0 & 33.3 & 20.2 & 46.4 & 0.0 & 33.3 & 0.0 & 33.3 \\
Constraint-aware & 84 & 100.0 & 17.9 & 9.5 & 72.6 & 0.0 & 17.9 & 0.0 & 17.9 \\
\bottomrule
\end{tabular}

}
\end{table}

This supplementary local table is useful mainly for one limitation: exact policy replay, static pinning, source quotas, and control-aware assembly preserve policy-sound carriage in these rows, yet unsafe-action proposals can remain nonzero because the model sometimes violates a preserved policy boundary. Table~\ref{tab:workflow-model-panel} reports model-level structured-output compliance and action-boundary outcomes for the same eight-assembler local triplet. The contrast between absent-state and preserved-unsafe buckets is the reason the workflow scaffold remains useful as a diagnostic: it separates assembly failures from model/action failures under preserved policy state without upgrading the supplement into reported action-level impact evidence.

\begin{table}[htb]
\centering
\small
\caption{Supplementary workflow-slice model panel over local Llama~3.1~8B, Qwen~2.5~7B, and Mistral~7B across eight assemblers and budgets 96/160/320/512. Values are percentages except $n$. Struct. counts parseable structured envelopes, including empty action envelopes; Text refuse counts safe refusals emitted as prose rather than as structured empty envelopes.}
\label{tab:workflow-model-panel}
\resizebox{\columnwidth}{!}{
\begin{tabular}{lrrrrrrrrr}
\toprule
Model & $n$ & Struct. & Text refuse & No action & Unsafe & Guarded exec. & Unguarded exec. & Absent-state & Preserved \\
\midrule
Llama 8B & 224 & 100.0 & 0.0 & 8.0 & 28.1 & 0.0 & 28.1 & 7.1 & 21.0 \\
Qwen 7B & 224 & 100.0 & 0.0 & 9.4 & 36.2 & 0.0 & 36.2 & 12.9 & 23.2 \\
Mistral 7B & 224 & 99.6 & 0.0 & 21.4 & 23.7 & 0.0 & 23.7 & 6.2 & 17.4 \\
\bottomrule
\end{tabular}

}
\end{table}

\paragraph{Pressure-only versus prompt-conflict factorial}
This exploratory local factorial separates pure workload pressure from explicit final-query conflict inside the supplementary diagnostic supplement. It uses the same seven workflows and local Llama~3.1~8B, Qwen~2.5~7B, and Mistral~7B models, but restricts the assembly treatments to hard truncation at budget 96 (policy evicted) and exact policy-last replay at budget 96 (policy visible). With neutral final queries, evicting the policy raises the instance-level unsafe-action proposal rate in this supplementary local slice, but the paired bootstrap interval crosses zero. Conflictive final queries are included only as a prompt-conflict stress condition, where policy visibility is not sufficient to ensure compliance. The local factorial therefore supports decomposition only; exact values remain in Table~\ref{tab:workflow-factorial-local} and the artifact, and the slice is not reported same-assembler action-level evidence or a strong pressure-only claim by itself.

\begin{table}[htb]
\centering
\small
\caption{Appendix-only paired local workflow factorial over seven workflows, three local paper models, budget 96, and two assembly treatments. $\Delta$ unsafe is hard truncation (policy evicted) minus exact policy-last replay (policy visible), paired by workflow, model, and final-query condition.}
\label{tab:workflow-factorial-local}
\resizebox{\columnwidth}{!}{
\begin{tabular}{lrrrrr}
\toprule
\textbf{Final query} & \textbf{Pairs} & \textbf{Visible unsafe} & \textbf{Evicted unsafe} & \textbf{$\Delta$ unsafe} & \textbf{95\% CI} \\
\midrule
Neutral & 21 & 28.6 & 38.1 & +9.5 & [-19.0, +33.3] \\
Conflict & 21 & 66.7 & 57.1 & -9.5 & [-33.3, +14.3] \\
\bottomrule
\end{tabular}

}
\end{table}

Table~\ref{tab:workflow-closed-model-panel} reports the closed-provider supplement after explicit cost estimation and approval, rescored offline with the reported action parser and extended with the same policy-last replay and recency-reminder baselines. The table is included as portability evidence for the supplementary workflow-slice scaffold, not as a headline claim. The deterministic guard blocks Policy-IR-expressible violations before execution; unsafe-action proposal and absent-state rates show what reaches the action boundary before enforcement. The low structured-output rate for Gemini~2.5~Flash indicates that this row is format-compliance limited under the reported action-envelope prompt.

\begin{table}[htb]
\centering
\small
\caption{Supplementary workflow-slice model panel over six closed models across six assemblers and budgets 96/160/320/512. Values are percentages except $n$. Struct. counts parseable structured envelopes, including empty action envelopes; Unsafe is the instance-level unsafe-action proposal rate before deterministic guard execution.}
\label{tab:workflow-closed-model-panel}
\resizebox{\columnwidth}{!}{
\begin{tabular}{lrrrrrrr}
\toprule
Model & $n$ & Struct. & Text refuse & Unsafe & Guarded exec. & Unguarded exec. & Absent-state \\
\midrule
anthropic-claude-haiku-4.5 & 168 & 100.0 & 0.0 & 6.5 & 0.0 & 6.5 & 2.4 \\
anthropic-claude-sonnet-4.6 & 168 & 100.0 & 0.0 & 19.6 & 0.0 & 19.6 & 14.3 \\
gemini-2.5-flash & 168 & 16.7 & 0.0 & 3.6 & 0.0 & 3.6 & 3.0 \\
gemini-2.5-flash-lite & 168 & 99.4 & 0.0 & 39.9 & 0.0 & 39.9 & 25.6 \\
openai-gpt-4.1-mini & 168 & 100.0 & 0.0 & 32.1 & 0.0 & 32.1 & 28.0 \\
openai-gpt-4o-mini & 168 & 100.0 & 0.0 & 49.4 & 0.0 & 49.4 & 32.1 \\
\bottomrule
\end{tabular}

}
\end{table}

Table~\ref{tab:workflow-closed-policy-panel} aggregates the same closed-provider supplement by assembler. In this supplemental run, exact policy-last replay preserves the policy boundary and has the lowest unsafe-action proposal rate, while a generic recency reminder does not carry the actual control boundary under pressure and behaves like ordinary lossy assembly.

\begin{table}[htb]
\centering
\small
\caption{Supplementary closed-provider workflow-slice results by assembler across six models and budgets 96/160/320/512. Values are percentages except $n$.}
\label{tab:workflow-closed-policy-panel}
\resizebox{\columnwidth}{!}{
\begin{tabular}{lrrrrrr}
\toprule
\textbf{Assembler} & \textbf{n} & \textbf{Eq.} & \textbf{Unsafe} & \textbf{Guard exec.} & \textbf{Unguarded exec.} & \textbf{Absent} \\
\midrule
Hard trunc. & 168 & 28.6 & 28.0 & 0.0 & 28.0 & 25.0 \\
Rolling summary & 168 & 28.6 & 32.1 & 0.0 & 32.1 & 28.6 \\
Hybrid & 168 & 28.6 & 27.4 & 0.0 & 27.4 & 23.8 \\
Policy-last replay & 168 & 100.0 & 13.7 & 0.0 & 13.7 & 0.0 \\
Recency reminder & 168 & 28.6 & 31.0 & 0.0 & 31.0 & 28.0 \\
Constraint-aware & 168 & 100.0 & 19.0 & 0.0 & 19.0 & 0.0 \\
\bottomrule
\end{tabular}

}
\end{table}

Table~\ref{tab:workflow-mechanism-audit} decomposes unsafe-action proposals by mechanism. The components sum to the instance-level unsafe-action proposal rate: \emph{Absent} means no applicable trusted control state remains visible in the decision state, \emph{Non-eq.} is the table label for visible control state that fails the exact-equivalence diagnostic, including security-unsound weakening and completeness-changing rewrites, and \emph{Preserved} means the policy boundary is preserved but the model still proposes an unsafe action. This separation is the operational reason to treat context assembly and deterministic guards as complementary rather than interchangeable.

\begin{table}[htb]
\centering
\small
\caption{Mechanistic audit of the local workflow-slice scaffold. Values are percentages except $n$; Unsafe is decomposed into absent, non-equivalent, and preserved-policy buckets.}
\label{tab:workflow-mechanism-audit}
\resizebox{\columnwidth}{!}{
\begin{tabular}{llrrrrr}
\toprule
Model & Assembler & $n$ & Unsafe & Absent & Non-eq. & Preserved \\
\midrule
Llama 8B & Hard trunc. & 28 & 21.4 & 14.3 & 0.0 & 7.1 \\
Llama 8B & Rolling & 28 & 14.3 & 7.1 & 0.0 & 7.1 \\
Llama 8B & Hybrid & 28 & 17.9 & 10.7 & 0.0 & 7.1 \\
Llama 8B & Policy replay & 28 & 28.6 & 0.0 & 0.0 & 28.6 \\
Llama 8B & Recency & 28 & 42.9 & 25.0 & 0.0 & 17.9 \\
Llama 8B & Static pin & 28 & 50.0 & 0.0 & 0.0 & 50.0 \\
Llama 8B & Source quota & 28 & 35.7 & 0.0 & 0.0 & 35.7 \\
Llama 8B & Control-aware & 28 & 14.3 & 0.0 & 0.0 & 14.3 \\
Qwen 7B & Hard trunc. & 28 & 21.4 & 17.9 & 0.0 & 3.6 \\
Qwen 7B & Rolling & 28 & 25.0 & 21.4 & 0.0 & 3.6 \\
Qwen 7B & Hybrid & 28 & 21.4 & 17.9 & 0.0 & 3.6 \\
Qwen 7B & Policy replay & 28 & 50.0 & 0.0 & 0.0 & 50.0 \\
Qwen 7B & Recency & 28 & 53.6 & 46.4 & 0.0 & 7.1 \\
Qwen 7B & Static pin & 28 & 53.6 & 0.0 & 0.0 & 53.6 \\
Qwen 7B & Source quota & 28 & 50.0 & 0.0 & 0.0 & 50.0 \\
Qwen 7B & Control-aware & 28 & 14.3 & 0.0 & 0.0 & 14.3 \\
Mistral 7B & Hard trunc. & 28 & 32.1 & 17.9 & 0.0 & 14.3 \\
Mistral 7B & Rolling & 28 & 25.0 & 10.7 & 0.0 & 14.3 \\
Mistral 7B & Hybrid & 28 & 28.6 & 14.3 & 0.0 & 14.3 \\
Mistral 7B & Policy replay & 28 & 25.0 & 0.0 & 0.0 & 25.0 \\
Mistral 7B & Recency & 28 & 21.4 & 7.1 & 0.0 & 14.3 \\
Mistral 7B & Static pin & 28 & 17.9 & 0.0 & 0.0 & 17.9 \\
Mistral 7B & Source quota & 28 & 14.3 & 0.0 & 0.0 & 14.3 \\
Mistral 7B & Control-aware & 28 & 25.0 & 0.0 & 0.0 & 25.0 \\
\bottomrule
\end{tabular}

}
\end{table}

Table~\ref{tab:workflow-local-threshold} shows why the expanded workflow supplement uses budget 96 as a critical pressure point. The fixture sweep uses the same seven workflows and unsafe action fixtures but no model calls. It isolates the assembler: below 512 tokens, ordinary truncation and summary policies often fail to carry trusted control state into the decision state, while exact policy-last replay, static policy pinning, source quotas, and control-aware assembly preserve the canonical policy boundary even at budget 96. At 512 tokens, ordinary truncation and summary policies also recover in this fixture, so the result should be read as a contested-budget threshold rather than evidence that larger effective budgets fail. A generic recency reminder does not carry the actual policy boundary under pressure, which separates ``remind the model to be careful'' from carrying the actual control state. The underlying artifact sweep continues through 1024/2048/4096/8192/16384 tokens; those no-overflow controls preserve the policy boundary for all assemblers and are omitted from the compact table because they add no further threshold transition.

\begin{table}[H]
\centering
\small
\caption{Expanded local workflow fixture threshold. Values are percentages; Eq. is exact equivalence and Absent is missing trusted control state.}
\label{tab:workflow-local-threshold}
\resizebox{\columnwidth}{!}{
\begin{tabular}{lrrrrrrrrrrrrrr}
\toprule
\textbf{Budget} & \multicolumn{2}{c}{\textbf{Hard trunc.}} & \multicolumn{2}{c}{\textbf{Rolling summary}} & \multicolumn{2}{c}{\textbf{Hybrid}} & \multicolumn{2}{c}{\textbf{Policy-last replay}} & \multicolumn{2}{c}{\textbf{Static pinning}} & \multicolumn{2}{c}{\textbf{Source quota}} & \multicolumn{2}{c}{\textbf{Control-aware}} \\
 & Eq. & Absent & Eq. & Absent & Eq. & Absent & Eq. & Absent & Eq. & Absent & Eq. & Absent & Eq. & Absent \\
\midrule
96 & 0.0 & 100.0 & 0.0 & 100.0 & 0.0 & 100.0 & 100.0 & 0.0 & 100.0 & 0.0 & 100.0 & 0.0 & 100.0 & 0.0 \\
160 & 0.0 & 100.0 & 0.0 & 100.0 & 0.0 & 100.0 & 100.0 & 0.0 & 100.0 & 0.0 & 100.0 & 0.0 & 100.0 & 0.0 \\
320 & 14.3 & 85.7 & 14.3 & 85.7 & 14.3 & 85.7 & 100.0 & 0.0 & 100.0 & 0.0 & 100.0 & 0.0 & 100.0 & 0.0 \\
512 & 100.0 & 0.0 & 100.0 & 0.0 & 100.0 & 0.0 & 100.0 & 0.0 & 100.0 & 0.0 & 100.0 & 0.0 & 100.0 & 0.0 \\
\bottomrule
\end{tabular}

}
\end{table}

\section{Reconstructed Agentic Decision-State Case}
\label{sec:appendix-agentic-case}
The artifact includes one reconstructed Qwen~2.5~7B pressure-plus-conflict case rebuilt from shipped records, with no live model call required. It combines workload pressure with a final query that asks to proceed before condition $Z$, so it is a decision-state visibility example, not a pressure-only causal isolate. Unmitigated truncation and B2+ expose zero applicable constraints and violate; B2+ plus SCP+ICE (A) preserves all three. Materials are under \path{artifact/docs/case_studies/agentic_case_study_20260318_seed11/}.

\begin{table}[H]
\centering
\small
\caption{Reconstructed Qwen~2.5~7B pressure-plus-conflict case.}
\label{tab:t7a-agentic-case}
\resizebox{\columnwidth}{!}{
\begin{tabular}{lccccc}
\toprule
\textbf{Arm} & \textbf{Viol.} & \textbf{Decision} & \textbf{Preserved} & \textbf{Visible} & \textbf{Ctx toks} \\
\midrule
Baseline truncation & yes & unsafe\_action & 0 & - & 260 \\
Structured summary & yes & unsafe\_action & 0 & - & 260 \\
Structured summary + SCP+ICE (A) & no & unclear & 3 & c1,c2,c3 & 260 \\
\bottomrule
\end{tabular}

}
\end{table}

\end{document}